\documentclass[sigconf]{acmart}

\AtBeginDocument{%
  }

\usepackage{color}
\usepackage{balance}
\usepackage{pdfpages}
\usepackage{enumitem}
\usepackage{float}
\usepackage{makecell}
\usepackage{longtable}

\setcopyright{acmlicensed}
\copyrightyear{2026}
\acmYear{2026}
\acmDOI{just_accepted}
\acmConference[CHIWORK '26]{CHIWORK '26: Proceedings of the 5th Annual Symposium on Human-Computer Interaction for Work}{June 22--25, 2026}{Linz, Austria}
\acmISBN{978-1-4503-XXXX-X/2018/06}

\begin{document}

\title{AI Disclosure with DAISY}


\author{Yoana Ahmetoglu}
\affiliation{%
  \institution{UCL Interaction Centre}
  \city{London}
  \country{United Kingdom}}
\email{yoana.ahmetoglu@ucl.ac.uk}

   \author{Marios Constantinides}
\affiliation{%
 \institution{CYENS Centre of Excellence}
 \city{Nicosia}
 \country{Cyprus}
 }
 \authornote{Also affiliated with University of Cyprus, Cyprus and University College London, UK.}
  \email{marios.constantinides@cyens.org.cy}

\author{Anna L. Cox}
\affiliation{%
  \institution{UCL Interaction Centre}
  \city{London}
  \country{United Kingdom}}
\email{anna.cox@ucl.ac.uk}

\renewcommand{\shortauthors}{Ahmetoglu et al.}

\begin{abstract}
The use of AI tools in research is becoming routine, alongside growing consensus that such use should be transparently disclosed. However, AI disclosure statements remain rare and inconsistent, with policies offering limited guidance and authors facing social, cognitive, and emotional barriers when reporting AI use. To explore how structured disclosure shapes what authors report and how they experience disclosure, we present DAISY (\underline{D}isclosure of \underline{A}\underline{I}-u\underline{S}e in \underline{Y}our Research), a form-based tool for generating AI disclosure statements. DAISY was developed from literature-derived requirements and co-design (\textit{N}=11), and deployed in a user study with authors (\textit{N}=31). DAISY-supported disclosures met more completeness criteria, offering clearer breakdowns of AI use across research and writing than unsupported disclosures. Surprisingly, despite concerns about how transparently disclosed AI use might be perceived, the use of DAISY did not reduce authors' comfort with the disclosure statements. We discuss design implications and a research agenda for AI disclosure as a sociotechnical practice. 
\end{abstract}

\begin{teaserfigure}
 \centering
 \includegraphics[width=1\textwidth]{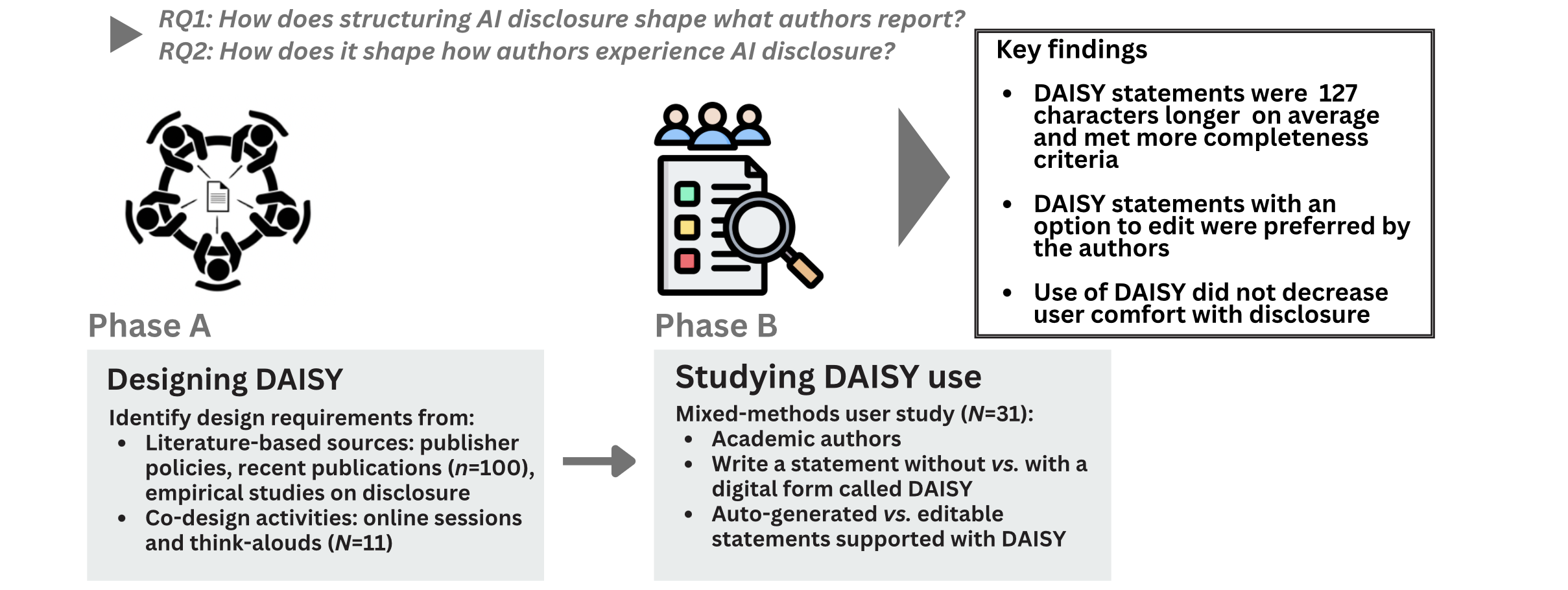}
 \caption{Overview of the design and user study of DAISY, a form-based tool for AI disclosure in academic writing, showing literature- and co-design–informed development (Phase A), a survey-based user study with authors (Phase B), and key findings.}
  \label{fig:teaser}
\end{teaserfigure}



\maketitle

\section{Introduction}
\label{sec:introduction}

A broad consensus has emerged in the scientific community that, when AI tools are incorporated into research, their use warrants disclosure \cite{council2024authorship, kaebnick2023editors, 10.1145/3711000}. For example, authors are expected to report when AI tools have been used to generate research ideas, draft or revise text, analyze data, or support methodological decisions. Such uses span a growing range of AI tools that are increasingly embedded in everyday research practice (e.g., large language models such as ChatGPT, code assistants like GitHub Copilot, and many others; for a review, see \cite{sontake2025review}). Disclosure of use of these kinds of AI tools is typically provided via a dedicated disclosure statement placed in the acknowledgements section, or an equivalent venue-mandated location \cite{ans2025presence}. Existing policies frame AI disclosure as supporting scientific norms of accountability and transparency by allowing others to assess how AI tools may have influenced the research and its reported claims \cite{resnik2025disclosing}.   

Despite broad agreement on the importance and purpose of AI disclosure in the ethics literature, AI disclosure statements remain largely absent in published research. A recent bibliometric study of medical education journals found that only around 2.5\% of articles included an AI-use disclosure statement \cite{ans2025presence}, even though survey evidence suggests that over 80\% of researchers use AI tools in at least some parts of their research and writing \cite{liaollms}. There may be several reasons for this gap. One possible reason is that disclosure can be challenging in practice. Existing policies require authors to disclose AI use beyond language-only edits but offer limited guidance on how such disclosures should be formulated (e.g., see the ACM policy on authorship and generative AI use \cite{ACMAuthorshipPolicy}). Another reason is that disclosure often involves social and reputational considerations; recent evidence suggests that declaring AI assistance can negatively influence how authors' competence and credibility are perceived \cite{reif2025evidence, brown2025academai}. A third possible reason is that disclosure relies on authors' ability to retrospectively (re)construct how AI tools were used throughout the research and writing process. Prior work shows that such (re)construction is often unreliable in mixed human-AI tasks, with individuals misattributing ideas or underreporting AI contributions~\cite{zindulka2025ai}. 

A structured disclosure format can reduce the effort involved in reporting AI use and make expectations clearer, which may help reduce underreporting. By offering a shared set of categories for common forms of AI assistance, such formats reduce the need for authors to invent wording or guess how their disclosures will be interpreted \cite{weaver2024artificial}. They may also help authors to reflect on where AI was used, helping to recall uses that might otherwise be omitted. Prior work in adjacent domains, such as model cards for machine learning systems \cite{mitchell2019model} and structured risk or impact reporting for AI technologies \cite{bogucka2025impact}, suggests that structured and standardized formats can effectively support disclosure. However, it remains unclear how these insights translate into the context of academic AI-use disclosure. While proposals for AI disclosure frameworks \cite{weaver2024artificial} and online AI label generators (e.g., \cite{AIUsageFacts2025}) have begun to emerge, these efforts have largely focused on proposing what could be disclosed rather than on empirically examining how structured supports shape reporting practices or how authors experience the task of producing structured AI-use disclosures in practice.

To investigate the questions of \textit{how structuring the AI disclosure process shapes what authors report}, and \textit{how they experience the task of disclosure for academic manuscripts}, we present DAISY (\underline{D}isclosure of \underline{A}\underline{I}-u\underline{S}e in \underline{Y}our Research), a digital AI disclosure form that guides authors through questions about their use of AI tools, and generates an AI disclosure statement for academic manuscripts (Figure~\ref{fig:teaser}). The tool was developed through elicitation of literature-based design requirements, and a co-design study with different academic stakeholders (\textit{N} = 11) and deployed in a mixed-methods user study with authors (\textit{N} = 31). DAISY produced longer and more complete disclosure statements, and participants preferred DAISY-generated statements (with the option to manually edit them) over statements written without DAISY. We also observed a disciplinary contrast. Non-HCI academics expected easy-to-use disclosure support that helps them navigate policies, while HCI academics imagined more complex logging and reflective systems focused on transparency and fairness that exceeded the current needs of most other academics. In so doing, we made three main contributions:
\begin{enumerate}[topsep=0pt]
    \item DAISY, a structured AI disclosure form that operationalises disclosure as an author-facing interaction (\S\ref{sec:daisy});
    \item empirical evidence of how structured AI disclosure reshapes both reported content and author experience, and how academics are currently using AI in their research practice (\S\ref{sec:evaluation});
    \item design implications and a research agenda for an ecology of AI disclosure tools that can serve diverse needs and disciplinary expectations (\S\ref{sec:discussion}).
\end{enumerate}

\section{Related Work}
\label{sec:related}

We surveyed various lines of research that our work draws upon, and grouped them into three main
areas: \emph{i)} research on AI disclosure spanning ethics and policy scholarship (\S\ref{subsec:rw1}); \emph{ii)} challenges on AI disclosure (\S\ref{subsec:rw2}); and \emph{iii)} design of documentation and reporting artefacts (\S\ref{subsec:rw3}).

\subsection{Policy and Ethics on Disclosing AI use in Academia}
\label{subsec:rw1}
We reviewed ethical analyses and publisher policies on AI-use disclosure in academia, which collectively define the goals and expectations of disclosure. We use this literature to surface a gap between normative guidance and the practical challenges authors face when constructing disclosure statements.

A recent 2024 survey with 816 researchers found that 81\% have started to use AI assistance with research activities in some capacity \cite{liaollms}. AI-assisted writing is also already appearing in publishing processes. Liang et al. \cite{liang2024monitoring} found widespread evidence of LLM-modified text in peer reviews at major AI conferences. Given such widespread adoption, it is increasingly accepted that AI-use in research must be transparent and responsibly reported.  Ethical analyses draw attention to the importance of AI-use disclosure for fairness, accountability, and scientific integrity. For example, Resnik and Hosseini \cite{resnik2025disclosing} argue that failing to disclose AI-use can distort the scientific record by masking who actually performed specific parts of the work. Similarly, Cohen and Moher \cite{cohen2025generative} note that inconsistencies in how authors report AI assistance make it difficult for editors, reviewers, and readers to assess the credibility of findings. Related concerns around attribution and responsibility in academic work have also been examined within HCI, for example in studies of how credit is assigned in multi-author publications \cite{early2018understanding}. At the same time, studies of academic writing and authorship practices suggest that norms of attribution and credit vary across disciplines and research communities \cite{cohen2025generative, early2018understanding}, implying that expectations around AI disclosure may likewise differ across contexts. Similar concerns about transparency, attribution, and ownership of AI-generated contributions have also been raised more broadly across creative and knowledge work domains \cite{epstein2023art}. One example is in journalism, where researchers have begun to examine practitioners’ expectations and needs around communicating AI involvement in news production \cite{venkatraj2025understanding}. This growing recognition is also visible within the HCI community, where recent discussions have indicated worrying uncertainty about when AI involvement should be disclosed and what form such disclosures should take \cite{10.1145/3707640.3729212}. 

In response to these concerns, publishers have begun to formalise expectations around AI disclosure. The ACM authorship policy, for example, permits the use of AI tools to assist with research activities but requires authors to fully disclose such use in their manuscripts, recommending explicit statements in the acknowledgments section \cite{ACMAuthorshipPolicy}. Comparable policies are being developed across publishers which focus on calls for transparency \cite{ElsevierPublishingEthics, SpringerNatureAuthorship}. However, existing policies leave substantial room for interpretation and limited direction on how authors should construct such statements, what level of detail is appropriate, or how responsibility should be expressed. They tend to frame AI disclosure as a reporting requirement rather than a practically challenging and cognitively demanding task. This overlooks the lived experience of constructing disclosure statements, including authors’ concerns about how their AI use is framed. Prior work on responsible AI has cautioned against reducing complex ethical practices to checklist-based compliance \cite{10.1145/3706598.3713184, 10.1145/3706598.3713278}. For example, Constantinides et al. \cite{10.1145/3686927} argue that RAI guidelines should function as interpretive and reflective instruments rather than signals of conformity. Applying this lens to AI disclosure suggests that disclosure statements should not be treated as a tick-box requirement, a concern also reflected in critiques of the “transparency ideal” as insufficient when divorced from practice \cite{ananny2018seeing}. 

\subsection{Challenges of AI Disclosure}
\label{subsec:rw2}
Creating AI disclosure statements in manuscripts is itself a complex reporting task. We discuss three types of AI disclosure challenges (i.e., \emph{social}, \emph{cognitive}, and \emph{emotional}) identified in prior HCI and related interdisciplinary work. While the salience of these challenges may vary depending on the purpose, level, and context of AI use, prior research consistently shows that uncertainty about how AI contributions should be interpreted, remembered, and evaluated can make disclosure difficult for authors \cite{brown2025academai, zindulka2025ai, chan2025understanding}. 


\subsubsection{Social Challenges}
A substantial body of work shows that declaring AI assistance can alter how one’s work and competence are judged by others. Reif et al. \cite{reif2025evidence} find that employees who use AI help are viewed as less competent, motivated, and hireable than those who rely on human assistance. In academic settings, Brown et al. \cite{brown2025academai} report widespread concern that peers, supervisors, or institutions may interpret AI use unfavourably, and that this anxiety is often amplified by inconsistent or ambiguous guidance. Reviewers likewise express suspicion toward AI-assisted manuscripts, raising concerns about fraud and diminished expertise \cite{hadan2024great}. More broadly, AI use has become entangled with moral stigma and perceptions of unfair advantage in knowledge work, with accusations of AI assistance functioning as a status-laden judgement about competence and legitimacy \cite{sarkar2025ai}. Such dynamics may lead individuals to fear being seen as less capable or less deserving of credit. This concern aligns with attribution research showing that when a salient external aid is present, observers tend to attribute outcomes to that aid rather than to the individual, systematically discounting the person’s effort and ability \cite{kelley1967attribution}.

\subsubsection{Cognitive Challenges}
Recent studies demonstrate that there are significant memory recall challenges when remembering and assessing AI contributions. While some of the studies were conducted in general writing or creative tasks rather than academic contexts, they provide relevant insights into how people remember and attribute AI contributions during human–AI collaboration. For example, Zindulka et al. \cite{zindulka2025ai} conducted an experiment in which participants completed creative writing and idea-generation tasks either alone, with AI assistance, or with mixed human-AI tasks. When tested one week later, participants showed substantial source-memory errors, often misattributing whether specific ideas or sentences had been written by themselves or by the AI, with the mixed human-AI condition producing the highest misattribution rates. Jakesch, Hancock, and Naaman \cite{jakesch2023human} show that people cannot reliably distinguish AI-generated self-presentations from human-written ones, and may be relying on intuitive but flawed heuristics that mislead their judgments. This difficulty extends to evaluating others’ use of AI. Nelson et al. \cite{nelson2025students} show that students overestimate teachers’ ability to detect AI-generated work, misjudging both detectability and the associated integrity risks. Skulmowski \cite{skulmowski2024placebo} suggested a theoretical account of why people misjudge their own contribution when assisted by AI, identifying two reasons: the \textit{AI placebo effect}, where assistance makes tasks feel easier and leads users to overestimate their own ability, and the \textit{ghostwriter effect,} where AI contributions are psychologically backgrounded, leading users to downplay or underreport the role of AI in the final output.

\subsubsection{Emotional Challenges}
Disclosing AI use can itself be emotionally challenging. The emerging construct of \textit{AI guilt} captures the moral discomfort some users feel when relying on AI tools for tasks traditionally seen as requiring human effort \cite{chan2025understanding}. The guilt is compounded when disclosure is incentivised by policy but social norms or audience perceptions remain uncertain. Indeed, authors who honestly disclose their AI use report shame or unease, reflecting a sense that honesty may be punished rather than rewarded \cite{bao2025ai}. Among students, higher levels of AI guilt are associated with reduced willingness to both use AI for creative tasks and openly report that use \cite{chan2025understanding}. In classroom contexts, ambiguous norms intensify anxiety and embarrassment when deciding whether to declare AI assistance \cite{aldulaijan2025impact}, and similar moral discomfort appears in other sensitive AI use settings \cite{luo2025seeking}.

\subsection{Design Interventions and Artefacts for Improving Disclosure Practices}
\label{subsec:rw3}
These challenges raise the question of how authors might be supported when reporting their AI-use in academic manuscripts. While relatively little design research has directly examined AI disclosure as an author-facing task, several adjacent strands of work have explored how structured disclosure artefacts can support accurate and compliant reporting while allowing authors discretion over how their work is represented.

Prior work on structured documentation for responsible AI introduced \textit{Datasheets for Datasets}, which propose a standardised format for documenting the motivation, composition, and intended use of datasets \cite{gebru2021datasheets}. Other work in machine learning documentation introduced \textit{model cards}, structured documentation artefacts that summarise a model’s intended use, evaluation procedures, performance characteristics, and limitations \cite{mitchell2019model}. Subsequent large-scale examinations show that many existing cards remain incomplete or omit critical sections \cite{raji2020closing}. Related approaches such as RiskRAG build on these templates by automatically pre-populating context-appropriate risk fields, aiming to increase reporting completeness \cite{rao2025riskrag}. This work aligns with prior HCI research showing that predefined disclosure templates can shape what information is revealed while reducing user effort in reporting \cite{gross2010disclosure}.

A separate line of research treats disclosure as a problem of information presentation, examining how AI-related information is structured and formatted to support comprehension by different stakeholders. Bogucka et al. \cite{bogucka2025impact} introduce \textit{Impact Assessment Cards} as a card-based format for communicating system-level information about AI systems, including their purpose, intended context of use, potential risks and benefits, and regulatory classification under the EU AI Act. This research argues for shifting disclosure away from long narrative reports toward a more compact format that supports quicker interpretation and comparison. Evaluations indicate that card-based designs support faster information extraction and higher-quality task performance than narrative documentation. These findings connect to earlier evidence that disclosure formats influence both the quantity and character of disclosure, for example, computer-administered forms tend to elicit greater self-report than face-to-face or paper formats due to reduced evaluation anxiety \cite{tourangeau2007sensitive}. 

To date, there is little peer-reviewed research that examines AI-use disclosure as an author-facing practice. One relevant contribution is the Artificial Intelligence Disclosure framework, which adapts contributorship taxonomies to the writing process by outlining stages such as conceptualisation, analysis, and drafting \cite{weaver2024artificial}. However, this work presents a conceptual structure rather than an evaluated disclosure tool. Alongside this, a small number of initiatives have proposed label-style disclosure formats inspired by nutrition labels, aiming to provide simple, standardised ways of listing common forms of AI assistance, though these efforts are not peer-reviewed and lack empirical evaluation ( e.g. \cite{AIUsageFacts2025}). Evidence from published research suggests that such proposals have so far had limited impact on practice: large-scale analyses show that AI-use disclosures remain rare and, when present, are typically brief and vague. For example, a review of over 2,000 medical education articles found that only 2.5\% included any AI-use disclosure, usually limited to naming the tool and asserting author responsibility, with little detail on how AI was used \cite{ans2025presence}.
\smallskip

\noindent\textbf{Research Gap.} Prior work has established the importance of transparency around AI use in academic publishing and identified a range of cognitive, social, and emotional factors that shape disclosure decisions \cite{reif2025evidence, brown2025academai, zindulka2025ai, skulmowski2024placebo}. In parallel, research on AI governance and machine-learning documentation shows that structured reporting formats can influence what information is recorded and how it is interpreted \cite{mitchell2019model, bogucka2025impact, rao2025riskrag}. However, this work has largely overlooked the moment when researchers are actually required to write AI disclosure statements for their own manuscripts. There is limited understanding of how authors interpret AI disclosure requirements, decide what to report, and experience the task of producing these statements in practice. Existing AI-use disclosures in published research remain rare and inconsistent \cite{ans2025presence, 10.1145/3613905.3650750}, and little empirical work has examined whether and how interactive tools can support authors in producing more complete and less burdensome disclosures \cite{AIUsageFacts2025, 10.1145/3707640.3729212, weaver2024artificial}. These gaps point to the need for research that empirically examines author-facing disclosure support tools and their effects on disclosure content and experience.

\section{DAISY}
\label{sec:daisy}

To examine how interactive tools might support the task of writing AI disclosure statements, we designed DAISY, a structured disclosure form intended to guide authors through reporting their AI-use (Figure~\ref{fig:teaser}). We followed a two-phase process to identify design requirements for the tool, adapting the approach of Bogucka et al. \cite{bogucka2025impact}. First, we derived design requirements by synthesising literature-based sources: publisher policies, analysis of disclosure statements from recent publications, and empirical research on how researchers use and disclose AI. Second, we used the initial design requirements as co-design materials to further develop them using a participatory approach. The aim of the co-design sessions was to contextualize the literature-based requirements by answering the questions of: \emph{i)} How do academics understand and reason about AI disclosure? and \emph{ii)} What forms of support for authors do academics envision for producing AI disclosures?

\subsection{Identify Design Requirements from Literature-based Sources and Co-Design Activities}

\subsubsection{Publisher Policies}
\label{subsecpolicy}
We reviewed publisher policies on AI-use in research. To scope the set of publishers included, we drew on a curated overview of publisher AI policies maintained by the UC Merced Library \cite{ucmerced-ai-policies}. At the time of writing, this overview provided one of the most widely used and consolidated summaries of publisher AI policies. Based on this overview, we included all 11 major publishers listed whose guidance academic authors are most likely to encounter, including ACM, IEEE, Springer Nature, Elsevier and Wiley, which together publish a substantial proportion of high-impact research across domains \cite{clarivate2023jcr}. For each publisher, we consulted and analysed the official policy documents available on the publisher’s website (see the policies in the \textit{Glossary} section of the study materials here: \cite{miro_coddesign_board}).

The review of the AI policies indicated that all publishers prohibit listing AI systems as authors. Minor assistive uses of AI for grammar, style, or basic editing typically do not require disclosure, whereas more substantive uses such as generating text, code, tables, figures, images, or conceptual content do. When disclosure is required, policies commonly ask authors to name the tool and version where relevant, indicate which parts of the manuscript or research process involved AI assistance, and describe the nature and purpose of that assistance. Publishers differ in the specificity of the guidance provided. Some specify where disclosures should appear, such as in a dedicated declaration, the Methods section when AI is used in the research process, or the Acknowledgements, while others leave placement to author judgement. Policies also differ in how explicitly they define disclosure content, from general guidance to more specific requirements with example wording. A small number include explicit restrictions, such as limits on AI-generated or AI-manipulated images, figures or research data. 

\subsubsection{Recent Publications}
To examine how AI-use disclosure is currently reported in practice, we conducted a targeted literature review of recently published research articles across a subset of major academic publishers \cite{pare2015synthesizing}. Using the set of publishers identified in (\S\ref{subsecpolicy}), we selected journals with reported acceptance rates below 30\% because acceptance rates below 30\% are commonly used in prior work to indicate selective publication venues (e.g. \cite{chen2010conference}). For each journal or venue, we sampled between 2 and 5 of the most recently published research articles (to achieve diversity within the publishers), resulting in a total sample of 100 papers. For each paper, we examined the acknowledgements, methods, and end matter for included AI disclosure statements. In addition, we conducted keyword searches within the PDFs using terms such as \textit{AI}, \textit{artificial intelligence}, \textit{ChatGPT} and \textit{Gemini}. These tools were included because prior surveys indicate that ChatGPT and Gemini are among the most commonly used AI tools in academic research \cite{liaollms, brown2025academai}. We found that only two papers in the sample contained explicit AI disclosure statements, which reported stylistic use of ChatGPT for language correction \cite{busboom2025tracing} or no use of AI tools at all. A list of all reviewed papers can be found in Appendix~\ref{sec:appendix-corpus-table}.

\begin{table*}[t] \centering \small \caption{Participant demographics and academic experience in the co-design study.} \setlength{\tabcolsep}{4pt} \renewcommand{\arraystretch}{1.1} \begin{tabular}{clllll} \hline \textbf{\#} & \textbf{Gender} & \textbf{Position} & \textbf{Field} & \textbf{Experience} & \textbf{Country} \\ \hline 1 & Female & PhD student & HCI & Author & UK \\ 2 & Male & Associate Prof. & Psychology & Author; Reviewer & UK \\ 3 & Female & PhD student & HCI & Author; Reviewer & UK \\ 4 & Male & Lecturer & Psychology & Author; Reviewer & UK \\ 5 & Male & Postdoctoral Res. & Mathematics & Author; Reviewer & UK \\ 6 & Male & Associate Prof. & Political Science & Author; Reviewer; Editor & UK \\ 7 & Female & PhD student & Education & Author; Reviewer & UK \\ 8 & Female & Professor & Sports \& Health & Author; Reviewer; Editor & UK \\ 9 & Male & Professor & HCI; Psych.; ML & Author; Reviewer; Chair; Editor & Austria \\ 10 & Female & Assistant Prof. & HCI \& CS & Author; Reviewer; Chair & Germany \\ 11 & Male & Lecturer & Machine Learning & Author; Reviewer & UK \\ \hline \end{tabular} \label{tab:demographics1} \end{table*}

\subsubsection{Empirical Studies}
Empirical studies cited in the Related Work section of this paper (\S\ref{sec:related}) were reviewed to understand how AI disclosure shapes authors’ social, emotional, and cognitive experiences, informing the design requirements presented in the following section, following the approach used by Bogucka et al. \cite{bogucka2025impact}. Similar to their work, we did not conduct a systematic literature review at this stage but used relevant prior studies as a starting point for identifying key challenges related to AI disclosure. These insights were subsequently contextualised through the co-design sessions and triangulated with publisher policy requirements and the analysis of existing disclosure statements. Prior work on social and reputational effects of AI disclosure informed requirements related to how disclosure is framed and experienced by authors, including concerns about judgement and stigma \cite{reif2025evidence, brown2025academai}, while studies on AI guilt and emotional discomfort informed requirements aimed at reducing emotional burden during disclosure \cite{chan2025understanding, bao2025ai}. Research on cognitive challenges in mixed human–AI contexts, including source-memory errors and misattribution of contributions, informed requirements aimed at supporting recall across the research and writing process \cite{zindulka2025ai, skulmowski2024placebo}. In this context, the Artificial Intelligence Disclosure framework provided an activity-level breakdown of research and writing tasks (e.g. ideation, analysis, drafting), informing which types of activities should be considered during disclosure \cite{weaver2024artificial}. Finally, work on structured disclosure and reporting artefacts informed requirements related to possible disclosure formats and interaction elements (e.g. forms, cards, labels), which were used as design ingredients in the co-design sessions (\S\ref{codesign}) \cite{tourangeau2007sensitive, gross2010disclosure, bogucka2025impact}.

\subsubsection{Synthesis of initial design requirements from literature-based sources}
We organised insights the literature-based sources into three groups of \emph{initial} design requirements that served as a starting point for the co-design sessions and, in some cases, directly informed the design of the disclosure artefact presented in this paper (as described in \S\ref{subsec:designing}).

Content requirements (R1) specified \textit{what} information a disclosure statement should include: disclosure of any meaningful use of generative AI beyond basic proofreading; naming the AI tool used and, where relevant, its version; indicating which parts of the research or manuscript involved AI assistance; describing the nature and extent of that assistance across research and writing activities, such as conceptualisation, methodology, information collection, data analysis, visualisation, writing, and administrative work; and affirming that authors retain full responsibility for the final content. Presentation requirements (R2) specified \textit{how} disclosure should be supported: providing guidance to support recall of AI involvement; using neutral phrasing to reduce shame or fear; generating ready-to-use disclosure statement. Context-of-use requirements (R3) specified \textit{when} and \textit{where} disclosure support should fit: enabling reuse across submissions and support for collaborative authorship; aligning with disclosure expectations across major publishers. 

\subsection{Conduct Co-Design Activities}
\label{codesign}

We conducted co-design activities to explore how academic stakeholders understand AI disclosure and what forms of support they envision to ground the literature-based requirements outlined above in people's lived experiences.

\subsubsection{Participants.}
We conducted 4 co-design sessions with a total of 11 academic authors, reviewers, academic journal editors and conference committee members (1-4 participants per session). This sample size is consistent with prior HCI co-design studies in related domains, which typically involve 10–15 participants in small groups (for example, \cite{bogucka2025impact}). We recruited participants through departmental mailing lists and personal networks, aiming for variation in career stage (doctoral researchers, postdoctoral researchers, faculty), disciplinary background, and familiarity with AI tools. Prior use of AI tools was not an explicit recruitment criterion for the co-design sessions. However, all participants reported having used AI tools in their research. Table \ref{tab:demographics1} provides demographic details. The study was approved by the UCL ethics committee.

\subsubsection{Procedure.}
Each session lasted 60 minutes and was conducted online with the help of a Miro board (which can be accessed here \cite{miro_coddesign_board}). Sessions were split in three parts: conversation, ideating and creating. In the conversation part, participants were introduced to a summary of publishers AI policies and to a brief overview of AI disclosure statements. They discussed if any of the information was surprising to them, and they shared their prior experience with AI disclosure policies and statements from their experience as authors, reviewers, conference chairs and editors where relevant. In the ideating part, participants were asked to imagine a tool that supports authors in creating an AI disclosure statement. They were invited to suggest requirements related to the content the tool should capture (R1), how this information should be presented (R2), and the context-of-use in which the tool would fit (R3), recording their ideas as sticky notes on separate boards for each participant which they later discussed with the group. The researcher noted collective requirements that were common in participants ideas. Next, participants were shown a new board presenting requirements derived from the literature and were invited to note which aspects they agreed or disagreed with. In the creating part, participants were asked to sketch how an AI disclosure artefact might look, using pen and paper to draw a tool over a 15-minute period, drawing on example elements (such as forms, cards, or agents), which they could get inspiration from. The design elements were based on prior work on card-based design elements to communicate AI risks \cite{bogucka2025impact}, form-based elements described in \cite{yell_form_elements}, and one chatbot-inspired design element based on a pilot study (P1), which was included in the main corpus. Participants took photos of their sketches and sent to the researchers via email. 

\subsubsection{Analysis.} 
\begin{figure*}
    \centering
    \includegraphics[width=1\linewidth]{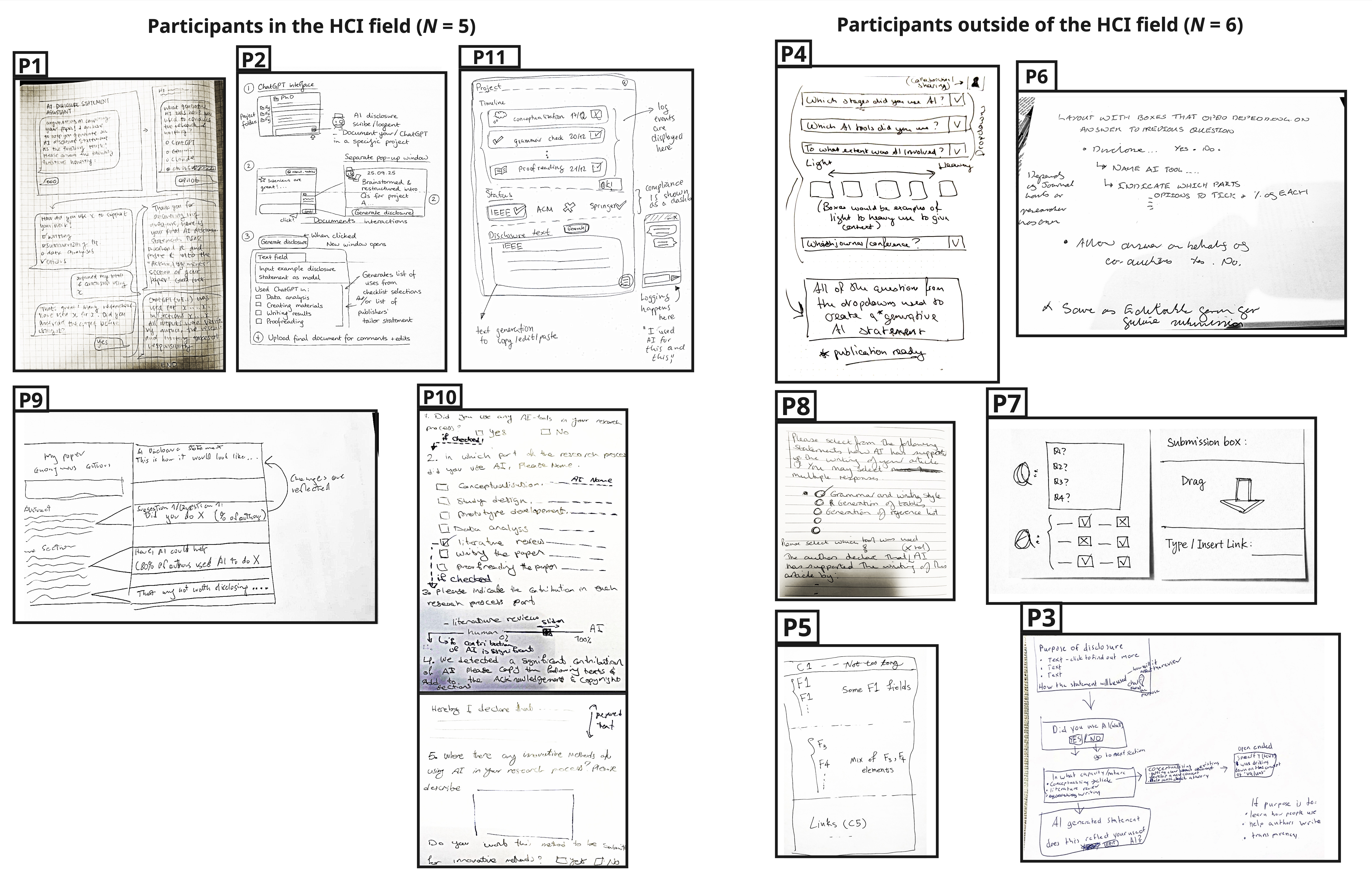}
    \caption{Speculative AI disclosure artefacts created by 11 participants (P1-11) during four co-design sessions, sorted by participant background and layout.}
    \label{fig:sketches}
\end{figure*}

All co-design sessions were recorded, transcribed, and analysed using inductive thematic analysis \cite{terry2017thematic}. The first author coded the transcripts and regularly presented the evolving codebook and emerging themes during research team discussions, where codes and theme definitions were iteratively refined using a shared Miro board. The sticky notes created during the second part of the co-design sessions were clustered based on the requirements they referred to, providing additional context for the analysis. Each sketch was supplemented by participants’ explanations and any relevant comments made during the co-design sessions (see Figure \ref{fig:sketches} for sketches, and Appendix \ref{sec:appendix-post-its} for post-its and Appendix \ref{sec:appendix-thematic} for a mind-map of initial thematic analysis used during the research group discussions). Data analysis meetings amounted to approximately four hours in total. 

\subsubsection{Results.}
The thematic analysis resulted in two main themes, each broadly corresponding to one of the research questions guiding the co-design activities about authors’ understandings of AI disclosure and their expectations of support for producing such disclosures.

\paragraph{Theme 1: Initial Sense-making and Questioning of AI Disclosure.}
Most co-design participants reported low prior level of engagement with AI disclosure policies, and the majority (8 out of 11) had never written an AI disclosure statement before. Awareness of disclosure requirements was often limited even among experienced reviewers and editors although with some exceptions (P9 and P10). In most cases, the discussions focused less on how AI disclosure should be implemented and more on why AI disclosure exists, whose interests it is intended to protect, and how it should function in the absence of clear norms or verification mechanisms.

\textit{Awareness.} Participants expressed uncertainty about what types of AI use should be disclosed. For example, language-correcting uses were often described as not something participants would think of declaring. P8 said: “Because I’m not asking it to create anything… I’m asking it to improve the language… it would never cross my mind to declare that”. However, during the discussion, this conclusion was reconsidered: the same participant later reflected on using AI to generate new pieces of text such as abstracts or paper titles, noting that “I have to admit I’ve done it for titles before… because I’m not very catchy or creative. Sometimes you want that punchier title, so you get it to come up with that” thereby recognising a form of contribution that went beyond language correction. 

\textit{Trust.} Participants often saw AI disclosure as operating as a trust-based “honour” system because there were no verification mechanisms. Editors and reviewers noted that they largely rely on authors’ self-reporting and have limited ability to assess whether disclosure statements fully capture how AI was used: “It’s purely the authors’ willingness to adhere to that disclosure, because there is no proof in any sense” (P11). While some participants saw this reliance on trust as consistent with broader academic norms, others (e.g. P9) expressed discomfort with how easily AI contributions could be minimised or selectively framed, particularly when minimal disclosures such as “used to improve grammar” could plausibly "mask" more substantive involvement. 

\textit{Purpose.} Several questioned the purpose of disclosure, for example, whether disclosure was intended to protect authors, inform reviewers or safeguard venues. P5 said: “I don’t understand what is the purpose of this… If someone doesn’t state it and uses AI, how do you catch them? Do you want to catch them?” Disclosure was often perceived as risky and encouraging strategic behaviour such as minimising detail or tailoring statements to venue-specific expectations. Other participants suggested that disclosure could also serve more constructive roles, such as supporting transparency or helping authors reflect on their own AI use. For example, P10 described encountering highly competent and creative uses of AI that they found “more admirable than judgable”, noting that such practices require “creativity” and “knowledge” but are not yet valued within academia: “the disclosure is always there looking like as if I’m policing someone… not the other way”. P10 suggested that disclosure mechanisms could instead make room for authors to describe how they work with AI, including innovative methods.

\paragraph{Theme 2: Tensions in Disclosure as a Protective Task or an Opportunity for Transparency}
Participants' sketches pointed to a tension between viewing AI disclosure as a protective and compliance-oriented task and as a potential opportunity for transparency about research practices. This tension was most strongly shaped by disciplinary background, particularly whether participants had experience in HCI, rather than by formal academic roles and editorial experience.

\textit{Disclosure as a Protective Task.} Most participants, especially those outside of the HCI field, sketched form-based interactions situated close to the point of manuscript submission. Disclosure was frequently framed as a means of meeting publisher requirements and protecting oneself from potential scrutiny. For these participants, disclosure had to be easy and minimally disruptive, with several explicitly noting that it “should not be another obstacle” (P5). Their sketches and written requirements sometimes located disclosure within the existing editorial systems, for example “in a form in the submission system” (P4). Several participants asked for guidance on what counts and what does not in terms of AI use for a statement, and for “a clear threshold” (P6) suggesting concern with drawing defensible boundaries. Others raised questions about verification, for example asking whether disclosure should “explain how it will be checked if you are being truthful or not” (P2).  
This framing was also visible in the sketches. P7's sketch showed a single submission box with a short checklist and a space to insert a disclosure text or link, while P4's showed a simple form starting with a yes/no question about AI use followed by fixed options indicating affected parts of the work. Several participants preferred to describe AI use in abstract terms, stating that they would “keep it high level” (P3) or “only say it if it was substantial” (P6), motivated by concerns about how disclosure would be perceived by the readers.

\begin{figure*}

    \centering
    \includegraphics[width=1\linewidth]{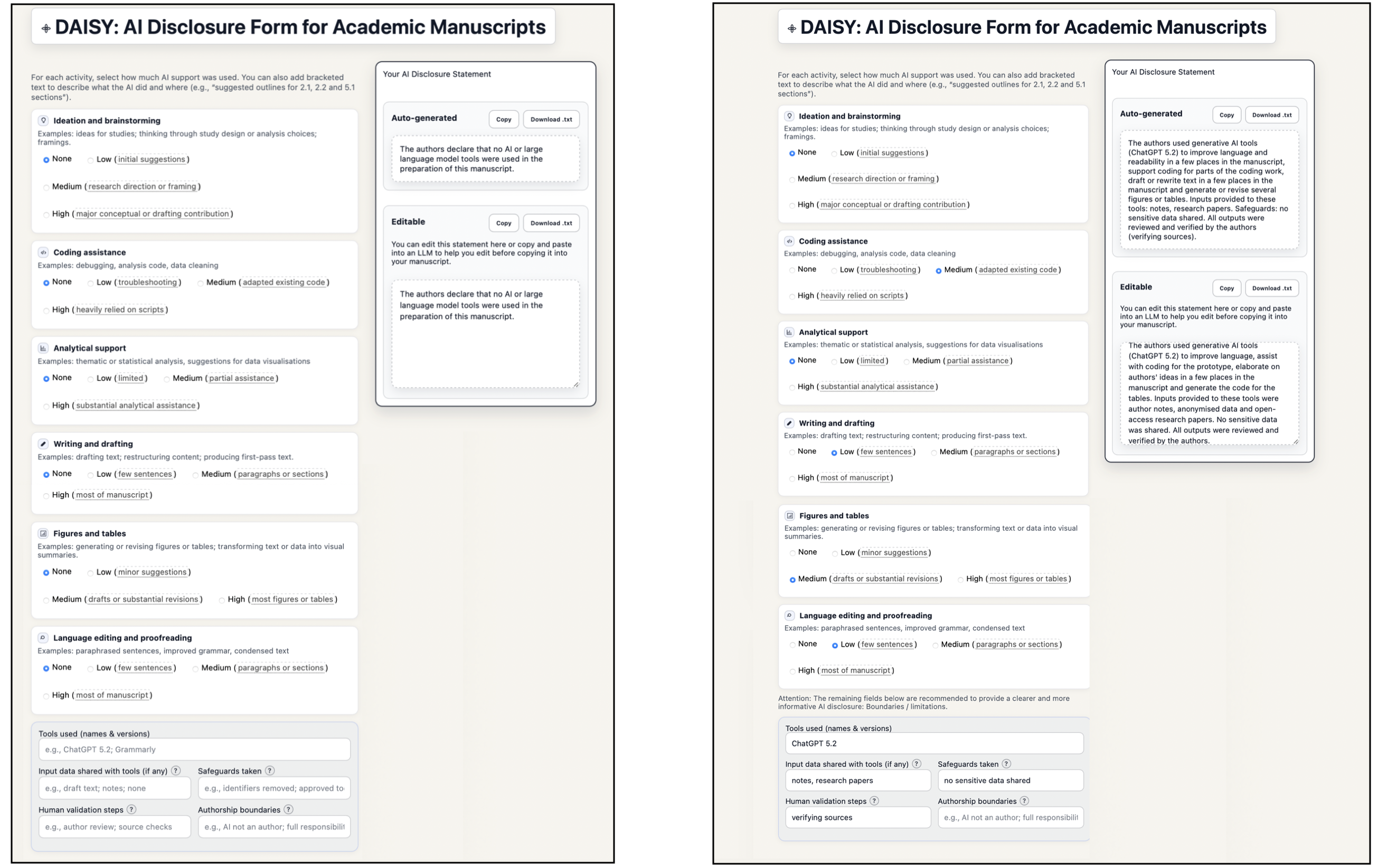}
    \caption{The DAISY interface. The left panel shows the default, unfilled form. The right panel shows a completed form and the resulting AI disclosure statement, including the editable output reviewed and refined by the author prior to submission.}
    \label{fig:daisy}
\end{figure*}

\textit{Disclosure as an Opportunity for Transparency.} The speculative disclosure artefacts were also described as a potential way to support transparency in research practices, a perspective raised primarily by participants with HCI backgrounds. Their sketches reflected concern with the complexity of AI involvement and with helping authors articulate how and where AI was used across research and writing activities, often aiming to support more detailed disclosure statements. A common concern was the difficulty of making sense of AI use after the fact, particularly when AI involvement was distributed across time, AI tools and research activities. Three out of five HCI participants proposed system-level tools rather than single forms. Some participants proposed logging approaches that allow AI use to be noted as it occurs and later summarised for disclosure (P9). In some cases, participants suggested checking these recorded interactions against the final manuscript to understand where and how AI may have shaped the work, with draft disclosure statements generated from this interaction history (P2). Others were more cautious about logging or quantification, questioning whether numerical representations of human-AI contribution meaningfully capture complex intellectual work, or whether such mechanisms risk shifting disclosure from a reflective practice to a monitoring one. P10 noted that disclosure should remain a “trust space”.


\subsection{Designing DAISY}
\label{subsec:designing}

To explore how a digital tool could support AI disclosure, we drew on the findings reported above and created DAISY. DAISY was designed to operationalise AI disclosure as an author-facing interaction and to enable empirical investigation of how authors construct and experience disclosure statements \cite{10.1145/1240624.1240704}. This decision to implement DAISY as a form-based interface reflects both existing disclosure mechanisms and the preferences expressed during the co-design sessions. More speculative approaches (e.g., design fiction) could explore alternative disclosure paradigms, but our aim was to study a format that could plausibly integrate with current practice. DAISY was built as a web application using HTML for structure, CSS for layout and styling, and JavaScript for interaction and statement generation, and is hosted on GitHub Pages.

To address content requirements (R1), DAISY structures disclosure around six activities: Ideation and brainstorming, Coding assistance, Analytical support, Writing and drafting, Figures and tables, and Language editing and proofreading. These categories were adapted from the AI disclosure framework but consolidated to meet the needs for brief and easy presentation (e.g. multiple writing-related subcategories are captured under Writing and drafting, and analysis-related uses under Analytical support) \cite{weaver2024artificial}. Authors indicate the level of AI support (none, low, medium, high) for each activity and can add short bracketed notes to specify what the AI did and where, allowing both high-level and more precise reporting in line with co-design findings. In addition to activities, DAISY includes open ended fields, such as tools used (including names and versions), input data shared with AI systems, safeguards applied during use, human review steps, and authorship boundaries indicating that responsibility for the final content remains with the authors. 

For presentation (R2), DAISY was implemented as a form-based interaction that generates a submission-ready disclosure statement and uses neutral language (e.g. reporting on activities as opposed to asking about judgments about how appropriate AI use was \cite{10.1145/3613905.3650750}). The form breaks disclosure into activity-based categories to support recall of categories that may be otherwise omitted \cite{zindulka2025ai}. DAISY outputs both an auto-generated statement and an editable version, allowing authors to adapt wording to fit their specific needs. In terms of context-of-use (R3), DAISY is designed for retrospective and submission-stage use rather than collaboration-heavy reuse across venues, reflecting that participants did not widely prioritise cross-submission reuse or co-author coordination in their accounts or sketches. 
\section{User Study}
\label{sec:evaluation}

Having designed DAISY, we examined authors’ experiences with it in a survey-based user study with academic authors (\textit{N} = 31). The study examined how structuring the AI disclosure process shapes both what authors report in their disclosure statements and how they experience the task of disclosure. To address these questions, we compared disclosure statements written without support, with DAISY’s auto-generated output, and with DAISY-supported editing in terms of statement length, disclosure completeness, overall preference and comfort with the resulting statements among other measures. The study also provides empirical insight into how academics currently use AI in their research and writing.  Below, we describe the study participants (§\ref{subsubsec:participants}), design (§\ref{subsubsec:design}), procedure (§\ref{subsubsec:procedure}), measures (§\ref{subsubsec:materials}), analysis (§\ref{analysis}), and the quantitative and qualitative findings (§\ref{subsec:results}). 

\subsection{Method}
\label{subsec:method}

\begin{table}[t]
\caption{Participant demographics in the evaluation of DAISY.}
\label{tab:demographics}
\centering
\small
\begin{tabular}{@{}p{0.30\linewidth}p{0.66\linewidth}@{}}
\toprule
\textbf{Characteristic} & \textbf{Summary} \\
\midrule
Gender & 17 women; 13 men; 1 non-binary \\
Age & 23--59 (\textit{M}=38, \textit{SD}=10) \\
Career stage &
10 PhD students/candidates; 7 postdoctoral researchers or research fellows; 4 assistant professors; 7 professors or associate professors; 3 other academic or research roles \\
Institutions \& geography &
20+ unique institutions (UK=10; US=4; others include France, Ireland, India, Netherlands, Australia, Switzerland, Brazil, Cyprus) \\
Field &
HCI=10; Computer Science=5; Environmental Sciences=5; Social Sciences=4; Engineering/Technology/Management=4; Psychology/Neuroscience=3 \\
\bottomrule
\end{tabular}
\end{table}

\subsubsection{Participants}
\label{subsubsec:participants}
Thirty-one participants were recruited via LinkedIn posts and campus flyers. These posts were shared through university channels with reach beyond the HCI community and the authors’ immediate networks, and contacts were encouraged to further circulate the call within their own academic networks. Table~\ref{tab:demographics} provides participant demographic information. None of the participants in the user study overlapped with those who participated in the earlier co-design or think-aloud sessions. Participants were required to be at least 18 years old and currently working in a research-related role. The study was approved by the UCL ethics committee. 

\subsubsection{Design}
\label{subsubsec:design}

We conducted a within-person mixed-methods study to evaluate the impact of DAISY on authors’ AI disclosure practices. To ground the task in a realistic context and maintain the ecological validity of a tool intended for real-world use, participants were asked to select an academic paper they were substantially involved in writing and practice writing AI disclosure statements for that paper under different conditions. This approach was chosen because producing an AI disclosure statement is tightly coupled to the content and history of a whole manuscript, which is difficult to reproduce meaningfully in a laboratory setting. The order of conditions was fixed: participants first wrote an AI disclosure statement on their own and then used DAISY. Although this introduces potential order effects, this approach was chosen for two reasons. Interacting with DAISY at the start would likely create strong order effects, as the tool presents structured disclosure questions and clarifications that go beyond the guidance most authors currently encounter, as observed in the co-design sessions. In addition, this sequence reflects an ecologically plausible scenario in which authors may first attempt to write a disclosure statement independently and only later encounter structured support tools such as DAISY. We discuss this methodological choice further in §\ref{limitations}.

\subsubsection{Procedure}
\label{subsubsec:procedure}

The study was conducted as a web-based survey using Qualtrics, with DAISY accessed through an external web interface. After providing informed consent, participants reported demographic information and were shown a brief definition of AI disclosure statements, without examples to avoid priming toward a particular structure or level of detail. Namely, the study displayed the following definition: ``An AI disclosure statement should explain whether and how AI tools were used, including which tools were used (and version), for what purposes (e.g., brainstorming, drafting, editing, analysis), and where in the paper they were applied, and state that responsibility for the content remains with the authors''. Participants then wrote a disclosure statement in an open-ended field. 

Next, participants interacted with DAISY, which generated an AI disclosure statement and provided a separate field for manual editing of the statement (Figure~\ref{fig:daisy}). They were instructed to write all disclosure statements as if submitting to a journal or conference manuscript. If no AI had been used for the selected paper, they were asked to state this explicitly. Based on the think-aloud sessions conducted during tool refinement (10–15 minutes), we estimate that completing the DAISY form itself typically takes approximately 5–10 minutes. Participants copied both the auto-generated DAISY statement and the edited version into the survey and completed the study measures. The survey concluded with background questions about prior AI use. 

\subsubsection{Measures}
\label{subsubsec:materials}

We measured the length and completeness of the disclosure statements produced under three conditions (no support, DAISY auto-generated, and DAISY edited). Length was operationalised as the number of characters in each statement. Completeness was assessed using a binary coding scheme indicating whether each disclosure statement included the following six elements, derived from publisher AI disclosure policies: (1) the name of the AI tool used; (2) the version of the tool; (3) the location of AI use within the manuscript; (4) the purpose of AI use (e.g., brainstorming, drafting, editing, analysis); (5) the extent or level of AI involvement; and (6) an explicit statement that responsibility for the content remains with the authors. 

To capture participants’ experience of using the resulting disclosure statements, we asked participants to rate how comfortable they would feel using each statement in a real manuscript on a Likert scale (0-10). We examined comfort as a subjective assessment of how at ease authors felt submitting the statement as written, because standardised support could plausibly either increase comfort (e.g., by reducing uncertainty or concerns about judgement) or reduce comfort (e.g., by constraining authorial control). Participants were also asked to indicate their overall preference among the three statements and to explain their choice in an open-ended response. We also collected self-reported measures of ease of use, likelihood of future use, and likelihood of recommending DAISY to colleagues, each measured on Likert scales (0-10).  Finally, we included open-ended questions to capture qualitative reflections on the disclosure process. Participants were asked whether their thinking about AI use changed while completing the form, whether any types of AI use were not captured by DAISY, and what improvements could be made to the tool.

\begin{table*}[t]
\centering
\small
\caption{AI-supported activities reported in disclosure statements (N = 31), with frequencies and illustrative examples.}
\label{tab:ai-activities-freq-quotes}
\setlength{\tabcolsep}{8pt}
\renewcommand{\arraystretch}{1.25}
\begin{tabular}{lccp{10cm}}
\hline
\textbf{AI-supported activity} & \textbf{Count} & \textbf{\%} & \textbf{Example disclosure excerpts} \\
\hline
Language and readability support 
& 25 & 80.6\% 
& ``used generative AI tools (ChatGPT 5.2) to improve language and readability in some parts of the manuscript''; 
``improve language (mostly grammar) across most of the manuscript'' \\

Drafting or rewriting text 
& 13 & 41.9\% 
& ``draft or rewrite text in some parts of the manuscript''; 
``rewrite text in a few places in the manuscript'' \\

Coding assistance 
& 12 & 38.7\% 
& ``support coding for small coding tasks like troubleshooting''; 
``support coding for a substantial portion of the coding work'' \\

Analysis support 
& 11 & 35.5\% 
& ``support analysis in a limited capacity by giving suggestions for data presentation''; 
``support analysis for parts of the work'' \\

Ideation and brainstorming 
& 6 & 19.4\% 
& ``support ideation and brainstorming occasionally by giving initial feedback on the author's prepared study idea and design''; \\

Figures or tables 
& 3 & 9.7\% 
& ``draft or refine figures and tables for a small number of figures or tables''; 
``generate or revise several figures or tables'' \\

Literature search or synthesis 
& 3 & 9.7\% 
& ``used to summarize reading notes''; 
``used to question literature''; \\

No AI use declared 
& 3 & 9.7\% 
& ``The authors declare that no AI or large language model tools were used in the preparation of this manuscript.'' \\
\hline
\end{tabular}
\end{table*}

\subsubsection{Analysis}
\label{analysis}

For the quantitative analysis, we compared disclosure statements produced under the three conditions using R. Statement length was analysed using non-parametric repeated-measures tests. Disclosure completeness was assessed using a policy-driven binary coding scheme indicating whether each statement included the disclosure elements (0–6) described in Measures~\ref{subsubsec:materials}. The coding was conducted by the leading author, who has domain expertise in AI disclosure research and training in qualitative coding. The leading author coded an initial subset (50\%) of the responses to develop preliminary results. These were then presented and discussed with the research team to refine category definitions and resolve ambiguities. The coding scheme was then finalised and applied to the remaining response. The resulting data were also analysed using non-parametric repeated-measures tests. Comfort ratings, preference rankings, and perceived ease of use were analysed descriptively. The open-ended survey responses were analysed using content analysis to identify recurring patterns and illustrative examples.

\subsection{Results}
\label{subsec:results}

\subsubsection{What Was Participants' Prior Experience with AI tools and AI Disclosure?}
Most respondents reported frequent AI use, with 18 indicating often and 7 very often, while only a small minority reported sometimes (2), rarely (3), or never (1) using AI tools to support their work. At the same time, most participants reported no prior experience writing AI disclosure statements themselves (16), while 6 had written one more than once and 5 had done so once, and in 4 cases the disclosure had been written by co-authors instead of the respondent. In their general research practice, most participants reported AI use for their research, with 15 describing the use of multiple tools and 13 mentioning a single or general tool, most commonly ChatGPT (17), followed by Gemini (7), NotebookLM (6), Perplexity (5), and Copilot (4), primarily for writing and language support (14), literature exploration and summarisation (10), information seeking and early-stage sensemaking (8), technical or coding support (6), and ideation or reflective support (5), while four participants reported limited or no prior AI use.

\subsubsection{What AI Uses did Participants Disclose with DAISY?} 

The disclosure statements produced as part of the study provide an additional window into how academics are using AI in their research practice. The most frequently reported use of AI in the disclosure statements was for improving language and readability, followed by drafting or rewriting text and support for coding, and analysis, with support for ideation and brainstorming, and figures and tables reported less often (Table~\ref{tab:ai-activities-freq-quotes} shows a breakdown of reported activities). It has to be noted that DAISY did not include a separate category for literature search or synthesis which was reported in three statements, therefore, indicating that the category was added during editing. In addition, ChatGPT was mentioned in 18 out of 31 statements (10 specifying ChatGPT 5.2), Grammarly in 7, Microsoft Copilot in 6, Google Gemini in 6, Claude in 3, Perplexity in 3, Google Labs in 1, and MAXQDA in 1. 

\subsubsection{Were DAISY-generated Statements Longer and More Complete?}
\begin{figure*}[t]
    \centering
    \includegraphics[width=0.48\linewidth]{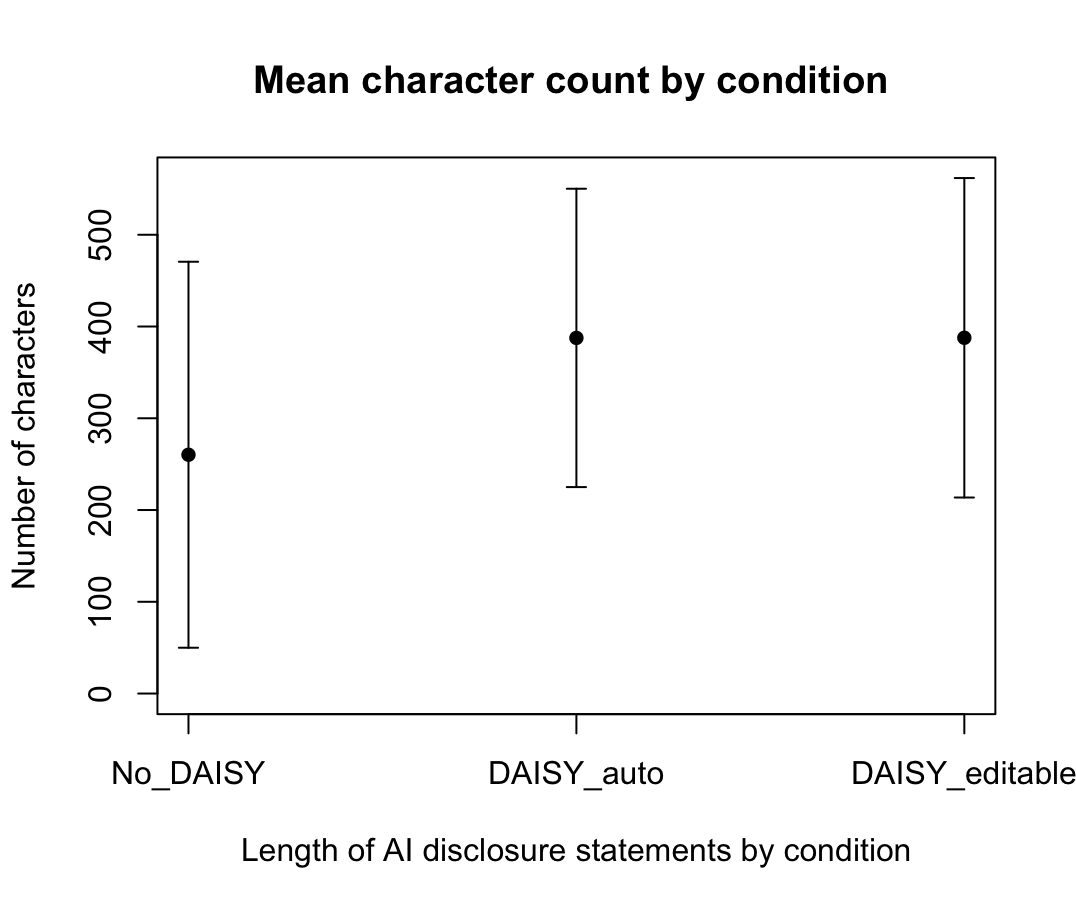}\hfill
    \includegraphics[width=0.48\linewidth]{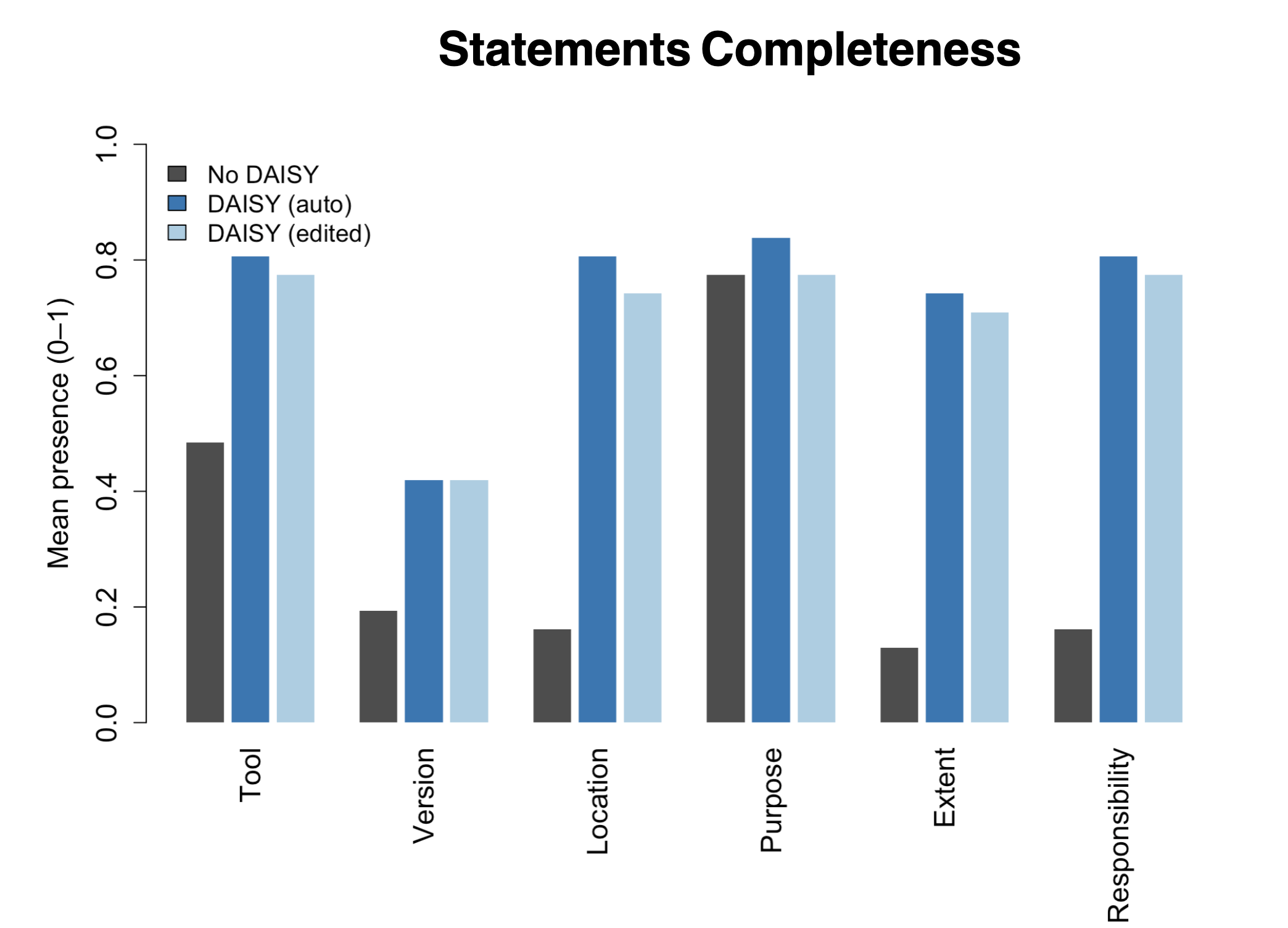}
    \caption{(Left) Mean character count in disclosure statements by condition. Error bars indicate $\pm$1 SD. (Right) Mean presence (0–1) of disclosure statements completeness criteria by condition.}
    \label{fig:wordcount-completeness}
\end{figure*}

Disclosure statements written with DAISY (auto-generated) (\textit{M} = 387.58, \textit{SD} = 162.58) and DAISY (editable) (\textit{M} = 387.71, \textit{SD} = 174.08) were significantly longer than those produced without DAISY (\textit{M} = 260.26, \textit{SD} = 210.33). A Friedman test indicated a significant effect of condition on character count, $\chi^2$(2) = 20.22, \textit{p} < .001, with a moderate effect size (Kendall’s \textit{W} = 0.33) (Figure~\ref{fig:wordcount-completeness}, left).

Statements produced with DAISY were, on average, more complete than statements written without tool support, for both auto-generated disclosures (\textit{M} = 4.42, \textit{SD} = 2.17) and edited disclosures (\textit{M} = 4.19, \textit{SD} = 2.40), compared to disclosures written without DAISY (\textit{M} = 1.90, \textit{SD} = 1.37). A Friedman test revealed a significant effect of condition on disclosure completeness, $\chi^2$(2) = 31.80, \textit{p} < .001. Post-hoc Wilcoxon signed-rank tests with Holm correction showed that disclosure statements written with DAISY were significantly more complete than those written without DAISY, both for auto-generated statements (\textit{p} < .001) and edited statements (\textit{p} < .001). There was no significant difference in completeness between auto-generated and edited DAISY statements (\textit{p} = .180).

\begin{figure*}
    \centering
    \includegraphics[width=0.95\linewidth]{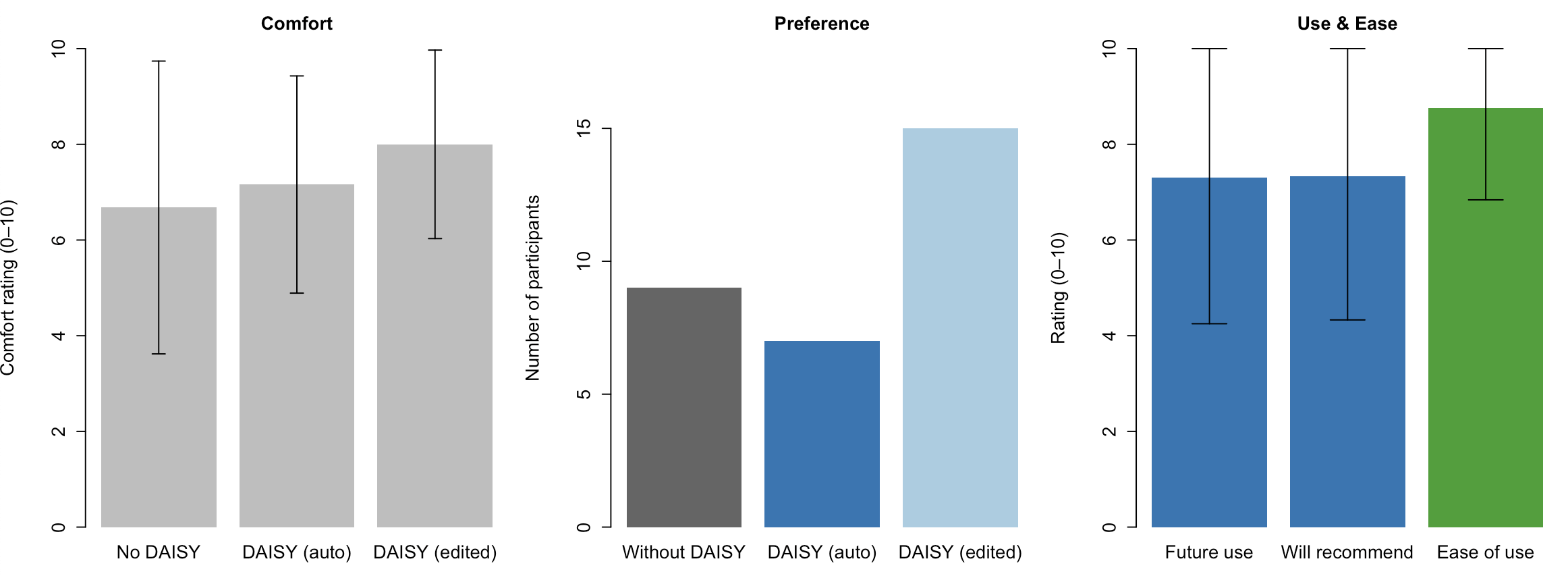}
    \caption{(Left) Mean comfort ratings (0–10) by condition (±1 SD). (Middle) Preferred disclosure approach, showing the number of
participants selecting each option. (Right) Participants’ reported likelihood of using DAISY in the future, recommending DAISY to
colleagues, and perceived ease of creating an AI disclosure statement with DAISY (0–10 scales; ±1 SD)}
    \label{fig:trio}
\end{figure*}

To identify which disclosure criteria showed the largest differences across conditions, we examined the means. The largest differences between DAISY and no-DAISY statements were observed for reporting the location of AI use within the manuscript, the extent of AI use, and authorial responsibility for the final content. These elements were rarely included in disclosures written without DAISY (Location: $M = 0.16$; Extent: $M = 0.13$; Responsibility: $M = 0.16$), but were present in the majority of DAISY-supported statements (auto-generated: Location $M = 0.81$, Extent $M = 0.74$, Responsibility $M = 0.81$; edited: Location $M = 0.74$, Extent $M = 0.71$, Responsibility $M = 0.77$); see Figure~\ref{fig:wordcount-completeness}. By contrast, criteria such as describing the purpose of AI use showed little difference across conditions (no-DAISY $M = 0.77$; auto $M = 0.84$; edited $M = 0.77$), while naming a specific AI tool (no-DAISY $M = 0.48$; auto $M = 0.81$; edited $M = 0.77$) and reporting its version (no-DAISY $M = 0.19$; auto $M = 0.42$; edited $M = 0.42$) showed more moderate increases with DAISY. 

To illustrate these effects, see the following disclosure statement written without DAISY: “We used ChatGPT (GPT-4o) to assist with grammar edits during manuscript preparation. All content was written and reviewed by the authors.” When supported by DAISY and editing the generated output, the statement was more detailed and the version of the tool used was revised: “The authors used generative AI tools (ChatGPT 5.2) to improve language and readability across most of the manuscript, draft or rewrite text in a few places in the manuscript and generate or revise several figures or tables. Inputs provided to these tools was draft text and safeguards included disabling training and retention. All outputs were reviewed and verified by the authors. AI is not an author.”

\subsubsection{What Were Participants' Preferences?} Participants reported relatively high comfort with using the statements in the real-world across all conditions. Mean comfort ratings increased from the no-DAISY condition (\textit{M} = 6.68, \textit{SD} = 3.06), to DAISY auto-generated statements (\textit{M} = 7.16, \textit{SD} = 2.27), and were highest for edited DAISY statements (\textit{M} = 8.00, \textit{SD} = 1.97). A Friedman test did not reveal a significant overall effect of condition on comfort ratings, $\chi^2$(2) = 3.98, \textit{p} = .137.
When asked which disclosure approach they preferred overall, nearly half of participants selected edited DAISY statements (15/31, 48.4\%). Fewer participants preferred writing disclosures without DAISY (9/31, 29.0\%) or using auto-generated DAISY statements (7/31, 22.6\%). Finally, participants reported generally positive evaluations of DAISY. Ratings (on a scale of 0-10) indicated high perceived ease of creating an AI disclosure statement with DAISY (\textit{M} = 8.76, \textit{SD} = 1.92). Participants also reported moderate likelihood of using DAISY in the future (\textit{M} = 7.31, \textit{SD} = 3.06) and in recommending DAISY to colleagues (\textit{M} = 7.34, \textit{SD} = 3.01) (Figure~\ref{fig:trio}).

\subsubsection{What was The Experience of Using DAISY?} We label survey participants as SP1–SP31 to distinguish them from participants in the co-design phase, and present the findings from the open-ended survey experience questions below. 

The findings showed that DAISY helped participants decide what to include in a disclosure and how to phrase it. SP5 wrote: ``DAISY helped me identify what to include in my disclaimer, as well as how to explain it''. The lack of external guidance was also raised: ``At the moment, there are no clear guidelines on how to report AI use. The one by DAISY seems to be the most genuine one'' (SP6). Similarly, participants said that completing the form improved how clearly they explained their AI use. For example, SP2 stated that ``my thinking did not change, but the way that I explained the process improved/became clearer''. Greater awareness of different forms of AI use was mentioned. SP11 wrote ``Yes, the DAISY form made me more aware of the different aspects AI could be used in a manuscript and helped me draft a more informative statement''. For some, the form prompted a reconsideration of what counts as AI use. SP8 wrote ``Completing the form prompted me to think more critically about the boundary between ‘standard’ digital tools and ‘AI assistance.’ Initially, I didn't think I needed a disclosure, but as I moved through the questions, I realized that my use of AI-driven translation tools and advanced grammar checkers actually constitutes a form of AI collaboration that warrants transparency''.

Several participants preferred to edit the generated text rather than use it as is. SP9 wrote: “I will combine my own version with the AI generated version. But still, my own will serve as the base”. Flexibility was valued: SP16 wrote “It allows for edits if needed, which offers flexibility”. Time savings were also mentioned: SP21 wrote “Easy to edit based on the autogenerated text, saving me lots of time”.
Participants differed in how they evaluated the auto-generated text. Some found it acceptable: “Auto-generated DAISY is perfectly fine” (SP24). Others criticised the language. SP25 wrote: “The auto-generated DAISY statement feels choppy and unprofessional” and SP17 wrote: “It sounds like a robot”. Further, SP23 wrote: “The language is nuanced and academics look for clues and meanings in language” and noted that phrases such as “few” could be misinterpreted by reviewers. Editing for clarity and safety was mentioned. SP22 wrote that they revised the text to make the scope of AI use “clearer and safer for authors to disclose”.

Participants also suggested changes to the form. Replacing open-ended text fields with predefined options was suggested. SP27 wrote: “Replace the open ended text fields with a few common standard choices (while leaving the option to customize)”. Clearer examples and task descriptions were also requested. SP29 wrote: “Provide clearer examples of AI disclosure statements for different research contexts”.
\section{Discussion}
\label{sec:discussion}

In this section, we first interpret the main empirical findings in relation to our research questions by focusing on how structured disclosure shapes both what authors report and how they experience the disclosure task. We then step back to consider the design implications of these findings, and use them to articulate a broader ecology of AI disclosure tools by reflecting on variation in authors' practices and disclosure goals. Finally, we outline limitations of our work and identify directions for future research.

Our findings should also be interpreted in light of the fact that AI disclosure norms in academic publishing are still emerging \cite{isc_ai_disclosure_standard}. While publisher policies and community discussions increasingly call for transparency about AI use, practices are still evolving and disclosure statements appear only infrequently in published work. This study therefore captures academic perspectives at a moment when expectations around AI disclosure are still being negotiated. Examining this early stage is important because the design of disclosure mechanisms may influence whether AI disclosure develops into a meaningful reflective practice or becomes a superficial compliance exercise \cite{10.1145/3613905.3650750}.

\subsection{Main Findings}

This section discusses the main findings of the user study in relation to our research questions. Specifically, we examine \emph{i)} how structuring AI disclosure shapes what authors report in disclosure statements, and \emph{ii)} how structuring AI disclosure shapes how authors experience the task of producing such statements. These findings suggest that form-based disclosure support can help authors produce more complete disclosures by encouraging the reporting of a wider range of AI uses and levels of use. 

\subsubsection{How does structuring AI disclosure shape what authors report?}
The findings show that structuring AI disclosure with DAISY led to longer and more complete statements than writing a disclosure without support, without reducing participants’ comfort with disclosure. As a result, DAISY-supported disclosures are likely more policy-aligned, in that they covered a greater number of reporting requirements \cite{ucmerced-ai-policies}. These improvements were mainly driven by helping authors report aspects of AI use that were often omitted otherwise, such as where AI was used in the manuscript, the extent of AI use, and authorial responsibility for the final content. At the same time, a minority participants preferred their own edited statements over the DAISY-generated versions, noting that their wording better conveyed limited AI use. This may point to the importance of anticipated interpretation by others in shaping authors’ comfort with AI disclosure and reveals a need for tools to allow authors to calibrate the tone and scope of their statements \cite{reif2025evidence}.

A key insight from these findings is that difficulties with AI disclosure may reflect uncertainty about what counts as AI use and how it should be reported, rather than reluctance to disclose. This interpretation is consistent with the co-design sessions, where participants often expressed limited awareness of what should be disclosed and uncertainty about whether routine or minor uses were relevant. It also aligns with prior work showing that people struggle to recall and attribute AI contributions, particularly in mixed human–AI tasks \cite{zindulka2025ai, skulmowski2024placebo}.

Only three participants added a new type of activity (literature-related support) when editing their statements, and no additional categories were suggested in the open-ended survey question. This limited expansion may reflect a broader tendency observed in HCI research, whereby users treat system-provided categories and framings as authoritative and do not actively challenge or extend them \cite{johnson2012beyond}. Alternatively, it may indicate that more condensed disclosure categories, such as those used in DAISY, are better suited to practice and more usable than more detailed disclosure frameworks proposed in prior work (e.g., the AID framework includes 14 categories \cite{weaver2024artificial}), particularly when the goal is to reduce reporting burden \cite{gross2010disclosure}. Future work could directly compare how the number and granularity of reporting categories shape what authors disclose about AI use.


\subsubsection{How does structuring AI disclosure shape how authors experience the task?}

Participants described the experience of using DAISY as making the task of writing an AI disclosure feel more structured and manageable. This experience of reduced cognitive effort is consistent with prior work showing that template- and form-based systems can offload decision-making and reduce uncertainty in sensitive reporting tasks \cite{gross2010disclosure}.

At the same time, while DAISY helped authors decide what to report, participants did not make many edits to the resulting statements themselves. Auto-generated text was often described as awkward, overly formal, or not sounding sufficiently natural, but manual editing, while useful, was also experienced as burdensome. Several participants indicated that, in practice, they would prefer to rely on their own familiar LLM tools to generate and polish disclosure statements, as these tools were already integrated into their writing practice and produced plausibly sounding AI disclosure statements. However, delegating disclosure writing to generative models introduces a potential limitation. Authors themselves may not always accurately recall or assess the full extent of AI assistance across a research process, and LLMs have no independent access to this history or criteria for judging the level of their contribution. As a result, disclosures generated in this way may appear coherent while still misrepresenting how much assistance was actually provided. When disclosure wording is generated separately from structured reporting, such omissions or distortions may be difficult to detect \cite{10.1145/3290605.3300233}.

As a result, participants’ experiences revealed a tension between support for deciding what to disclose and the effort involved in shaping how the disclosure is worded. This tension reflects a broader pattern identified in prior research on human–AI collaboration. Rather than avoiding AI involvement altogether, people tend to prefer systems that support decision-making while allowing them to retain control over the intent and implications of the final output, particularly in tasks that are socially evaluated \cite{jakesch2023human}. In such contexts, users often welcome assistance with structure or ideation, while remaining sensitive to how final outputs are phrased and interpreted. In the case of AI disclosure, this sensitivity is amplified by evidence that disclosing AI use can negatively affect perceptions of competence and legitimacy \cite{reif2025evidence, chan2025understanding}.

We also observed disciplinary differences in how participants responded to DAISY depending on their background. Participants with an HCI background more often suggested additional and more complex features, whereas participants from non-HCI disciplines tended to prefer simpler forms of support that required little effort and were primarily focused on compliance or risk reduction. One possible explanation is that HCI researchers are more accustomed to reflecting on their interactions with AI during everyday work and to engaging with speculative or exploratory tool designs \cite{rogers2012hci}. This difference may also reflect a form of design bias, where HCI participants imagine sophisticated tools that align with research ideals but may be less likely to be adopted in practice. Conversely, non-HCI participants’ preference for simpler support may reflect closer alignment with real-world constraints and institutional pressures. While we cannot determine the reasons for these differences, they point to the importance of considering disciplinary context when designing and evaluating AI disclosure tools. 

\subsection{Implications}

While our findings raise a range of potential implications, we focus here on a small set of design implications that are most directly supported by our empirical results and relevant to the design of disclosure support tools like DAISY.

Disclosure tools may benefit from a strong assumption that authors are accountable for the text and therefore may need to avoid presenting generated disclosures as final. Instead, they may prioritise editability to support authors in producing statements they can stand behind. Therefore, disclosure tools could treat the identification of discloseable AI use and the linguistic formulation of a disclosure as distinct design problems. This implication aligns with long-standing mixed-initiative design principles in HCI, which argue that systems should support users in making judgements while reserving automation for downstream execution tasks (e.g., phrasing or formatting), rather than allowing automation to implicitly make decisions on the user’s behalf (e.g., Principles of Mixed-Initiative User Interfaces \cite{10.1145/302979.303030}). More recent HCI work on AI-assisted writing similarly adopts workflows in which users first specify constraints or intent, and only then use generative models to produce or refine text within those boundaries (e.g., PromptChainer \cite{10.1145/3491101.3519729}; Luminate \cite{10.1145/3613904.3642400}). Applying this pattern to AI disclosure suggests that tools should first support authors in reasoning about what forms of AI use are relevant to report, before offering any language generation constrained by those decisions. This separation can be implemented in practice through multiple output styles. A minimal output reports AI use at a high level and avoids detailed descriptions of extent or location unless explicitly required. More specific outputs incorporate additional details provided by the author in situations where these are deemed appropriate.

Further, given that existing AI disclosure policies provide limited guidance on how disclosures should be written, disclosure tools could focus on translating high-level requirements into concrete activity-based structures that authors can work with. In practice, this means offering templates and prompts that make expected disclosure elements visible and actionable, while allowing authors flexibility in how they express them. Our findings suggest that publishers could play an important role in enabling and evaluating such disclosure support. Because authors are typically motivated to engage with disclosure only when it is required, publisher submission processes provide a context in which disclosure tools can be meaningfully trialled and studied. Embedding tools like DAISY at these points may offer opportunities to support authors while also enabling empirical evaluation of disclosure practices at scale. This may be especially useful given ongoing efforts to establish more standardised expectations for AI disclosure wording \cite{isc_ai_disclosure_standard}.

\subsection{Toward an Ecology of AI Disclosure Tools}

\begin{figure}
    \centering
    \includegraphics[width=0.99\linewidth]{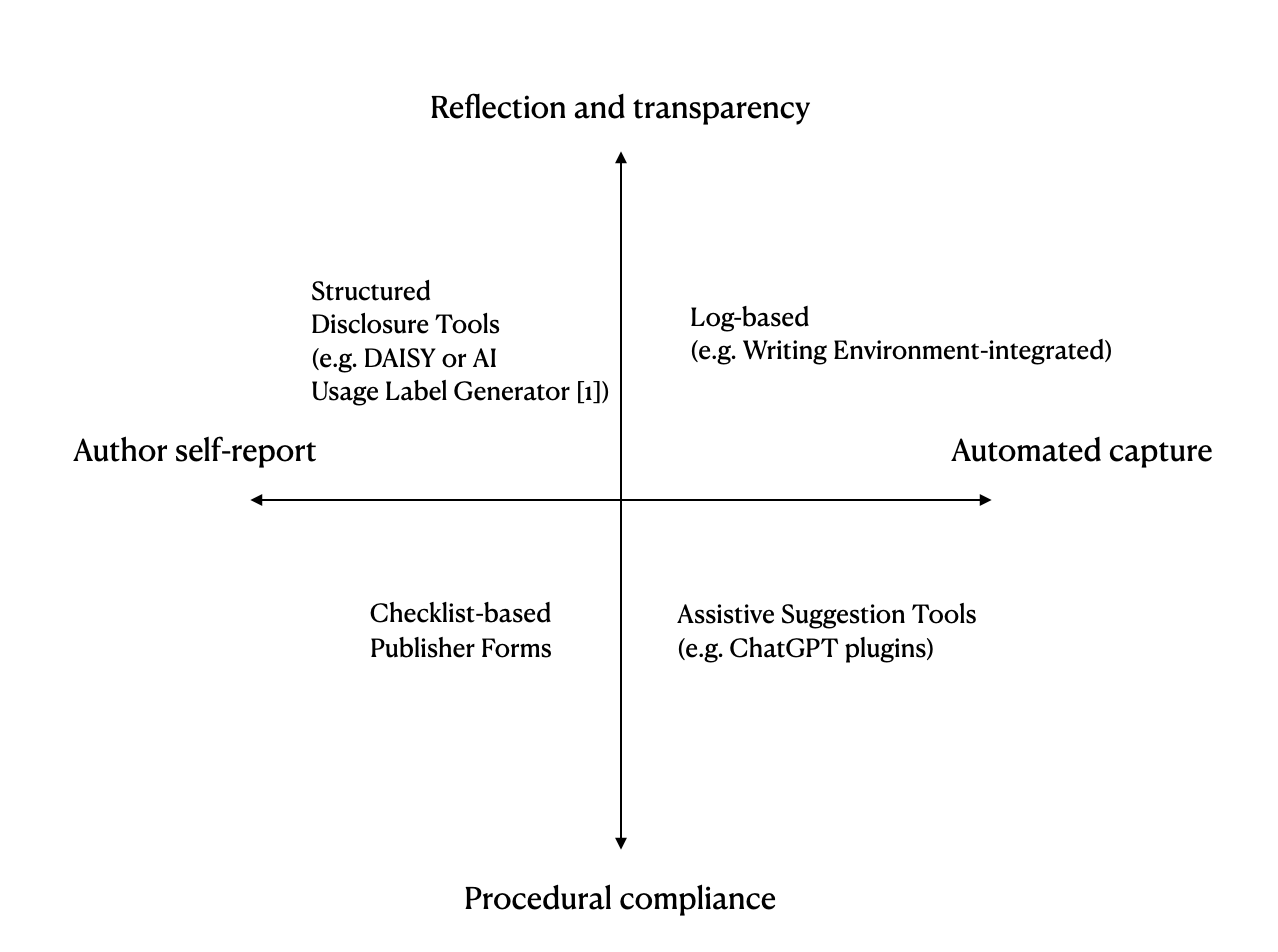}
    \caption{A speculative conceptual design space for AI disclosure tools. The horizontal axis contrasts author self-report with automated capture, while the vertical axis contrasts procedural compliance with reflection and transparency. The four quadrants illustrate distinct approaches to disclosure identified through our findings.}
    \label{fig:ecology}
\end{figure}

Our co-design and survey findings indicate that AI disclosure is not a single tool design problem. Participants differed in their AI practices, disciplinary norms, tolerance for effort, and expectations of disclosure, and often described tensions between compliance, transparency, and control. As a result, no single form of disclosure support was seen as appropriate across contexts. To reflect this diversity, we present a speculative conceptual design space that outlines several possible approaches to supporting AI disclosure.

Figure~\ref{fig:ecology} presents a speculative conceptual design space that maps both existing and potential AI disclosure tools along two dimensions identified as important in our study: the extent to which disclosure relies on author self-report versus automated capture, and whether disclosure primarily serves procedural compliance or supports reflection and transparency. 

The lower-left quadrant represents checklist-based publisher forms. These tools rely on author self-report and are oriented toward procedural compliance. They minimise effort by asking authors to confirm AI use in simple or binary ways, but offer limited support for reflection on how AI shaped the work. Several participants described such tools during the co-design activities. The upper-left quadrant captures structured disclosure tools, such as DAISY or previously suggest AI usage nutrition label-style generator \cite{AIUsageFacts2025}. These tools also rely on author self-report, but are designed to support reflection and transparency rather than simple compliance. By prompting authors to reason about different forms of AI use and articulate them in structured ways, these tools foreground author judgement while encouraging more meaningful disclosure.

The lower-right quadrant includes assistive suggestion tools, such as general-purpose LLM plugins that generate disclosure text. These tools reduce effort by automating phrasing, but typically rely on limited inputs and provide little support for reflection or accountability. As a result, they may streamline disclosure writing while leaving key judgements about what should be disclosed underspecified. Our study provides evidence that this quadrant naturally exists in authors practice. The upper-right quadrant represents log-based systems that rely on automated capture to support high levels of transparency. By recording interactions with AI tools directly, these systems can reduce reliance on memory or self-report and enable detailed reconstruction of AI use. However, they also raise challenges around surveillance, privacy, and authorial control, making them more suitable for contexts with strong transparency requirements or institutional support. While some academics with editorial and review experience expressed a need for such tools, their design and use would warrant careful study.

\subsection{Limitations and Future Work}
\label{limitations}

This study was not intended to define or validate a comprehensive taxonomy of AI use for disclosure. DAISY therefore uses a small set of high-level prompts. Given the structured nature of the form, it is not surprising that DAISY-supported disclosures covered more categories of AI use. Our contribution was not to show that structure increases reporting, but to examine how this additional content was perceived by authors and whether they chose to remove or revise it. Participants rarely edited disclosures to remove categories, suggesting that structured prompts can surface discloseable AI use without reducing comfort or authorial control. These findings provide initial insight into how disclosure support may shape emerging AI reporting practices. Further, 

This study examined a retrospective disclosure task in an online survey, capturing authors’ perspectives at a single point in time rather than long-term use or integration into real submission workflows. In addition, many participants had backgrounds in HCI, which, although not reflected in our quantitative results, may have shaped their familiarity with or comfort in reflective disclosure tasks. Future work should examine disclosure tools in real-world submission contexts, across disciplines, and over time, and further investigate how different disclosure formats are interpreted by reviewers, editors, and readers. More automated or tightly integrated disclosure approaches also warrant careful study, particularly with respect to privacy, accuracy, and governance.

Future work could explore more controlled experimental designs, such as counterbalanced condition orders or between-subjects comparisons, to better isolate potential order effects while still accounting for the strong learning effects introduced by structured disclosure tools like DAISY. Further, the sample of participants is skewed toward HCI and related computing fields (15 out of 31). Future studies could also deliberately oversample from disciplines less familiar with AI tool use and disclosure norms to examine whether the observed patterns hold across fields.

Moreover, although the completeness score was derived from a systematic review of publisher AI policies, it represents an operationalisation developed by the authors rather than an externally validated measure of policy compliance. Therefore, higher scores should be interpreted as indicating that disclosures include more of the policy-derived reporting elements used in this study, rather than definitive compliance with publisher requirements. Future research should seek external validation from journal editors, publishers, or policy experts to assess whether disclosures scoring higher on these criteria would be considered adequate in practice. The use of example values in the interface (e.g., “ChatGPT 5.2”) may have influenced participants’ responses by signalling the expected level of specificity. These examples were included because participants reported difficulty imagining what should be written in AI disclosure statements and indicated that concrete examples would make the tool easier to use. Future studies could compare interfaces with and without examples to examine their effect on disclosure completeness and wording. Lastly, although our co-design findings suggested disciplinary differences in how participants conceptualised disclosure tools, the user study was not powered to statistically examine disciplinary differences. Future work could compare researchers from different fields.

Looking ahead, structured disclosure tools could also function as research instruments in their own right. By capturing disclosure-relevant information in a consistent manner, tools such as DAISY could enable longitudinal and comparative studies of how authors understand, negotiate, and report their use of AI over time and across contexts. Such data could support future work examining where uncertainty or ambiguity in disclosure persists, how norms around acceptable AI use evolve, and how disclosure practices differ across disciplines, publication venues, or stages of the research process. Future studies could also compare authors’ DAISY-supported disclosures with independently logged records of AI tool use, where available, to examine the alignment between reported and observed practices and to better understand the limits and accuracy of self-reported AI disclosure. More broadly, this line of work positions AI disclosure as part of a developing human–AI work ecology, in which expectations around responsibility, accountability, and authorship are likely to continue shifting as generative AI becomes further embedded in academic practice \cite{10.1145/3729176.3729202}.
\section{Conclusion}
\label{sec:conclusion}

As AI tools become increasingly embedded in academic research, expectations around disclosure are expanding faster than the support available to authors. This work examined how structured form-based disclosure support shapes both what authors report and how they experience the task of producing disclosure statements. Our findings show that structured disclosure support can help authors produce more complete disclosures, especially when authors retain control over how their AI use is expressed. Rather than pointing to a single optimal solution, this work argues for understanding AI disclosure as an ecology of tools that serve different purposes across contexts. This positioning brings into focus AI disclosure as an evolving sociotechnical practice requiring further study.

\section{AI Disclosure Statement}
The authors used generative AI tools (ChatGPT 5.2) to support language improvements across the manuscript, to assist with the generation and revision of tables (Tables~\ref{tab:demographics1}, \ref{tab:demographics}, and \ref{tab:ai-activities-freq-quotes}) based on author-provided instructions, to provide limited analytical inspiration (e.g., helping refine interpretations already developed by the authors), to assist with writing and debugging R code used for data analysis, and to support parts of the implementation of the DAISY tool (e.g., drafting code under author direction). All AI-generated outputs were reviewed and verified by the authors, who take full responsibility for the final content. The use of generative AI followed relevant ethical guidance.
This disclosure statement was created using DAISY (available at: \url{https://dryoanaahmetoglu.github.io/daisy-disclosure/})

\begin{acks}

We thank all study participants for their time and contributions.

\end{acks}

\balance

\bibliographystyle{ACM-Reference-Format}
\bibliography{main}

\newpage

\appendix
\onecolumn
\section*{Appendix}
\label{sec:appendix-corpus}

\section{Corpus of Analyzed Publications}
\label{sec:appendix-corpus-table}

\begin{longtable}{p{0.9cm} p{3.0cm} p{1.2cm} p{9.0cm}}
\caption{Corpus of analyzed publications included in this study.}
\label{tab:corpus} \\
\hline
\textbf{ID} & \textbf{Venue / Publisher} & \textbf{Year} & \textbf{Title} \\
\hline
\endfirsthead

\hline
\textbf{ID} & \textbf{Venue / Publisher} & \textbf{Year} & \textbf{Title} \\
\hline
\endhead

C1 & ACM CHI & 2025 & ``It's impressive, but in practice...'': Experiencing a Realistic Digital Transformation in and beyond the Classroom \\
C2 & ACM CHI & 2025 & A Stakeholder Value Framework for Augmentative and Alternative Communication \\
C3 & ACM CHI & 2025 & ``Did you sleep well?'': A Multimodal Sleep Diary for Sustained Self-Reporting by Children \\
C4 & ACM CHI & 2025 & ``All Day, Every Day, Listening to Trauma'': Investigating Features of Digital Interventions for Empathy-Based Stress and Burnout \\
C5 & ACM CHI & 2025 & AVEC: An Assessment of Visual Encoding Ability in Visualization Construction \\

W1 & CHIWORK & 2025 & Framing the (in)visible: Insights into Visibility Practices of Remote Knowledge Workers \\
W2 & CHIWORK & 2025 & ``ChatGPT, Don't Tell Me What to Do'': Designing AI for Context Analysis in Humanitarian Frontline Negotiations \\
W3 & CHIWORK & 2025 & When Efficiency Meets Fulfillment: Understanding Long-Term LLM Integration in Knowledge Work \\
W4 & CHIWORK & 2025 & ``Being a nanny isn't just caregiving'': An Analysis of How Nannies Seek Support in Online Communities like/r/Nanny \\
W5 & CHIWORK & 2025 & Tracing Transformations of the Modern Workplace and Imagining its Future \\

E1 & Elsevier & 2026 & Recurrent U-Net-based Graph Neural Network (RUGNN) for accurate deformation predictions in sheet material forming \\
E2 & Elsevier & 2025 & A Systematic Review on the Influence of Feeding Expressed Mother’s Own Milk Using Varying Expression Practices or Treatments on Health and Growth of Recipient Infants \\
E3 & Elsevier & 2025 & Urban city of tomorrow: Bicycle based measurements of particulate matter exposure before and after road closure to motorized traffic \\
E4 & Elsevier & 2025 & Gamified Feedback in Adaptive Retrieval Practice: Points and Progress-Bars Enhance Motivation but not Learning \\
E5 & Elsevier & 2025 & The Trusted Partner for Financial Decision Making: Romantic Partner or AI? \\
E6 & Elsevier & 2025 & User Personas, Ideation and Large Language Models: A Post-Hoc Study \\
E7 & Elsevier & 2025 & Toward efficient vibe coding: An LLM-based agent for low-code software development \\
E8 & Elsevier & 2025 & From humans to algorithms: A sociotechnical framework of workplace surveillance \\
E9 & Elsevier & 2026 & Doing regions: multiplicity and singularization in the ontological politics of the Arctic \\
E10 & Elsevier & 2026 & Effects of urban internal and external migration on urban settlement, commuting, and employment containment in Melbourne, Australia \\

M1 & Emerald & 2025 & The unaccounted effects of digital transformation: implications for accounting, auditing and accountability research \\
M2 & Emerald & 2025 & Leader--employee perfectionism (in)congruence and role clarity: a role theory approach to employee workplace well-being \\
M3 & Emerald & 2025 & AI and learning experiences of international students studying in the UK: an exploratory case study \\
M4 & Emerald & 2025 & A machine learning model to detect early math content in YouTube videos \\
M5 & Emerald & 2025 & How do cultural values affect economic growth? An empirical evidence from world values survey (1994--2021) \\
M6 & Emerald & 2024 & An investigation on immigration inflows, GDP productivity and knowledge production in selected OECD countries: A panel model analysis \\
M7 & Emerald & 2025 & Generative AI in construction risk management: a bibliometric analysis of the associated benefits and risks \\
M8 & Emerald & 2025 & Assessing environmental quality and health implications of slaughterhouses’ operation within urban residential settings of a developing country \\
M9 & Emerald & 2025 & Advancing health equity amongst displaced persons through telehealth services: a retrospective cross-sectional study \\
M10 & Emerald & 2026 & Is a shorter working week in the construction industry smart and sustainable? \\

I1 & IEEE & 2024 & Machine-Learning-Assisted Optimization for Antenna Geometry Design \\
I2 & IEEE & 2025 & Rapid Deployment and Semi-Autonomous Retrieval of a Tracked Robot Using an Unmanned Transport Vehicle for Post-Disaster Exploration \\
I3 & IEEE & 2025 & From Intent to Accountability: Exploring the Role of Mental States in Robot Accountability \\
I4 & IEEE & 2025 & Transforming Electricity Safety Training and Culture using active methodologies \\
I5 & Oxford Academic & 2023 & Quantifying the Cost of Web Accessibility Barriers for Blind Users \\

S1 & Springer Nature & 2026 & Role of behaviour change in controlling the 2022 Paris mpox outbreak \\
S2 & Springer Nature & 2026 & Frequent presymptomatic household transmission of influenza A but not influenza B virus \\
S3 & Springer Nature & 2026 & School-based suicide prevention using the gatekeeper programme: a cluster-randomized trial \\
S4 & Springer Nature & 2026 & A simplified wearable device powered by a generative EMG network for hand-gesture recognition and gait prediction \\
S5 & Springer Nature & 2026 & Body-induced electroluminescence for bio-inspired 3D spatial position perception \\
S6 & Springer Nature & 2026 & Wearable lateral flow assays for cortisol monitoring with time-dynamic sweat sampling and sensing by electrochromic timers \\
S7 & Springer Nature & 2026 & A noise-tolerant human--machine interface based on deep learning-enhanced wearable sensors \\
S8 & Springer Nature & 2026 & An X-ray-emitting protocluster at $z \approx 5.7$ reveals rapid structure growth \\
S9 & Springer Nature & 2026 & A Cambrian soft-bodied biota after the first Phanerozoic mass extinction \\
S10 & Springer Nature & 2026 & Advancing regulatory variant effect prediction with AlphaGenome \\

O1 & Oxford Academic & 2026 & The Cultural Applicability of Desistance Research within the Arab (Syrian) Region \\
O2 & Oxford Academic & 2026 & A Zemiology of Torturous Violence and the Limits of Law \\
O3 & Oxford Academic & 2026 & Fetal Positions: Understanding Cross-National Public Opinion about Abortion \\
O4 & Oxford Academic & 2026 & Power, Prejudice and Opportunism: Tracing Vigilante Crimes and Social Harm in India \\
O5 & Oxford Academic & 2026 & Deciding how many judges should decide: The question of constitution benches at the Indian Supreme Court \\
O6 & Oxford Academic & 2026 & The Dispersion of Power. A Critical Realist Theory of Democracy \\
O7 & Oxford Academic & 2026 & Pluralizing federalisms \\
O8 & Oxford Academic & 2026 & The Collaborative Constitution \\
O9 & Oxford Academic & 2026 & Dignity and Judicial Authority \\
O10 & Oxford Academic & 2025 & Explicating miRNA-mediated regulation of inflammatory pathways in COPD, MS, and lung cancer using explainable artificial intelligence: insights from peripheral blood profiles \\

P1 & PLOS & 2026 & Resveratrol inhibits bladder cancer proliferation by targeting the AURKA/STAT3 axis: From computational analysis to experimental validation \\
P2 & PLOS & 2026 & Improving infection prevention and control in Ghana primary health care facilities: Evaluation of the STREAM disinfectant generator \\
P3 & PLOS & 2026 & Prevalence and mechanisms of high-level carbapenem antibiotic tolerance in clinical isolates of \textit{Klebsiella pneumoniae} \\
P4 & PLOS & 2026 & \textit{Metarhizium anisopliae}: A fungal biocontrol agent against \textit{Rhipicephalus microplus} \\
P5 & PLOS & 2026 & YAP inhibits HIV-1 transcription and promotes HIV-1 latency by regulating E3 ubiquitin ligase UHRF1 mediated tat degradation \\
P6 & PLOS & 2026 & HIV impairs and exploits pulmonary Th17 and Th22 cell-mediated immune responses to \textit{Mycobacterium tuberculosis} \\
P7 & PLOS & 2026 & Correction: Predicting unsafe behaviour from the objective assessment of fatigue manifestation among scaffolders: Evidence from a quasi-experimental simulation study \\
P8 & PLOS & 2026 & Factors associated with unmet healthcare needs in patients using primary care access points for unattached patients in Quebec (Canada) \\
P9 & PLOS & 2026 & Use of Kaplan--Meier and Cox regressions in the distribution of length of stay in animal shelters for pre-specified calendar periods: Definition, computation, and examples of dog length of stay in Orange County, California \\
P10 & PLOS & 2026 & Correction: Metagenomic characterization of bacterial abundance and diversity in potato cyst nematode suppressive and conducive potato rhizosphere \\
P11 & PLOS & 2026 & Fast identification of the charging pile plug materials using laser-induced breakdown spectroscopy \\
P12 & PLOS & 2026 & Correction: Open science practices in a Portuguese nursing higher education institution: An exploratory, descriptive study \\
P13 & PLOS & 2026 & Correction: Absence of ultimate controller and investment efficiency: Evidence from China \\
P14 & PLOS & 2026 & Mental health help-seeking among individuals with breast cancer: A qualitative exploration of women’s and healthcare practitioners’ perspectives \\

G1 & SAGE & 2026 & Finding compromise and cooperation through PAR for spiritual rituals in the Magaliesberg Biosphere Reserve, South Africa \\
G2 & SAGE & 2026 & Women to women research for economic empowerment in Uganda: A feminist participatory action research project \\
G3 & SAGE & 2025 & Culturing developmental friendship for a politics of love \\
G4 & SAGE & 2025 & Centering Indigenous knowledges in community-based participatory research \\
G5 & SAGE & 2026 & A practical guide for integrating community-engaged research across the psychological research cycle \\
G6 & SAGE & 2026 & CATAcode: A principled approach for coding check-all-that-apply demographic items \\
G7 & SAGE & 2025 & Language models accurately infer correlations between psychological items and scales from text alone \\
G8 & SAGE & 2025 & Investigating the barriers and enablers to data-sharing behaviors: A qualitative registered report \\
G9 & SAGE & 2026 & Human papillomavirus disease prevention in the United States and Germany among gender diverse adults with a cervix \\
G10 & SAGE & 2026 & Violence and associated risk factors among Hijra and transgender persons in India: Analysis of the integrated biological and behavioral surveillance survey \\

T1 & Taylor \& Francis & 2026 & Transnational citizenship: A study of citizenship identity among the Dayak Bidayuh in the Indonesia--Malaysia border region \\
T2 & Taylor \& Francis & 2026 & Citizenship in the undeportation regime: Unpredictable and situated inclusion at the borders of Europe \\
T3 & Taylor \& Francis & 2026 & Independentist narrative among diasporan Rojhelatî (Eastern) Kurds: Practices of transborder citizenship, dynamics and lines of contestation \\
T4 & Taylor \& Francis & 2025 & The reluctant citizen: The nation, state and the citizen in India \\
T5 & Taylor \& Francis & 2025 & Contested beings: Citizenship along the riparian zone in the India--Bangladesh borders \\
T6 & Taylor \& Francis & 2025 & Transnational citizenship pathways of Chinese professional migrants in Singapore: Negotiating provisionality, permanence, and emerging precarity in (post-)pandemic times \\
T7 & Taylor \& Francis & 2025 & Contested citizenship in the liminal spaces of a divided Cyprus \\
T8 & Taylor \& Francis & 2025 & Working on the farm: Acts of citizenship and accommodative resistance \\
T9 & Taylor \& Francis & 2025 & Return of liminality: Singapore Malay-Muslim return migrants from Australia \\
T10 & Taylor \& Francis & 2025 & The Spanish deportation regime: Unpacking hierarchies of (non)citizenship through the lens of deportabilization \\

Wl1 & Wiley & 2026 & The marriage of metal nanoclusters with reticular frameworks: Synthetic strategies and biomedical applications \\
Wl2 & Wiley & 2026 & Placental crises: Disruptive selection and maternal under-investment as the foundations of mammalian placental evolution and dysfunction \\
Wl3 & Wiley & 2026 & The power of many: When genetics met yeasts and high-throughput \\
Wl4 & Wiley & 2026 & Metabolite profiling and anticancer evaluation of Iraqi \textit{Haloxylon articulatum} halophyte via LC--HRMS/MS, molecular docking, and ADMET analysis \\
Wl5 & Wiley & 2026 & Four new triterpenoids from \textit{Rubus pirifolius} Smith \\
Wl6 & Wiley & 2026 & Natural flood algorithm for efficient parameter identification in a proton exchange membrane fuel cell models \\
Wl7 & Wiley & 2026 & Recent advances toward greener synthesis of quinolines and benzophenanthridines \\
Wl8 & Wiley & 2026 & Boc$_2$O-promoted cascade reactions of thioepoxides: Unexpected formation of oxazolidinones featuring a C5 sulfur-modified side chain \\
Wl9 & Wiley & 2026 & Design, synthesis, and biological evaluation of C-15 arylated sclareol derivatives against inflammation: In vitro and in silico studies \\

\hline
\end{longtable}

\newpage
\section{Post-It Notes Created During the Co-Design Sessions} 
\label{sec:appendix-post-its}

\begin{figure*}[h!] 
    \centering 
    \includegraphics[width=0.7\linewidth]{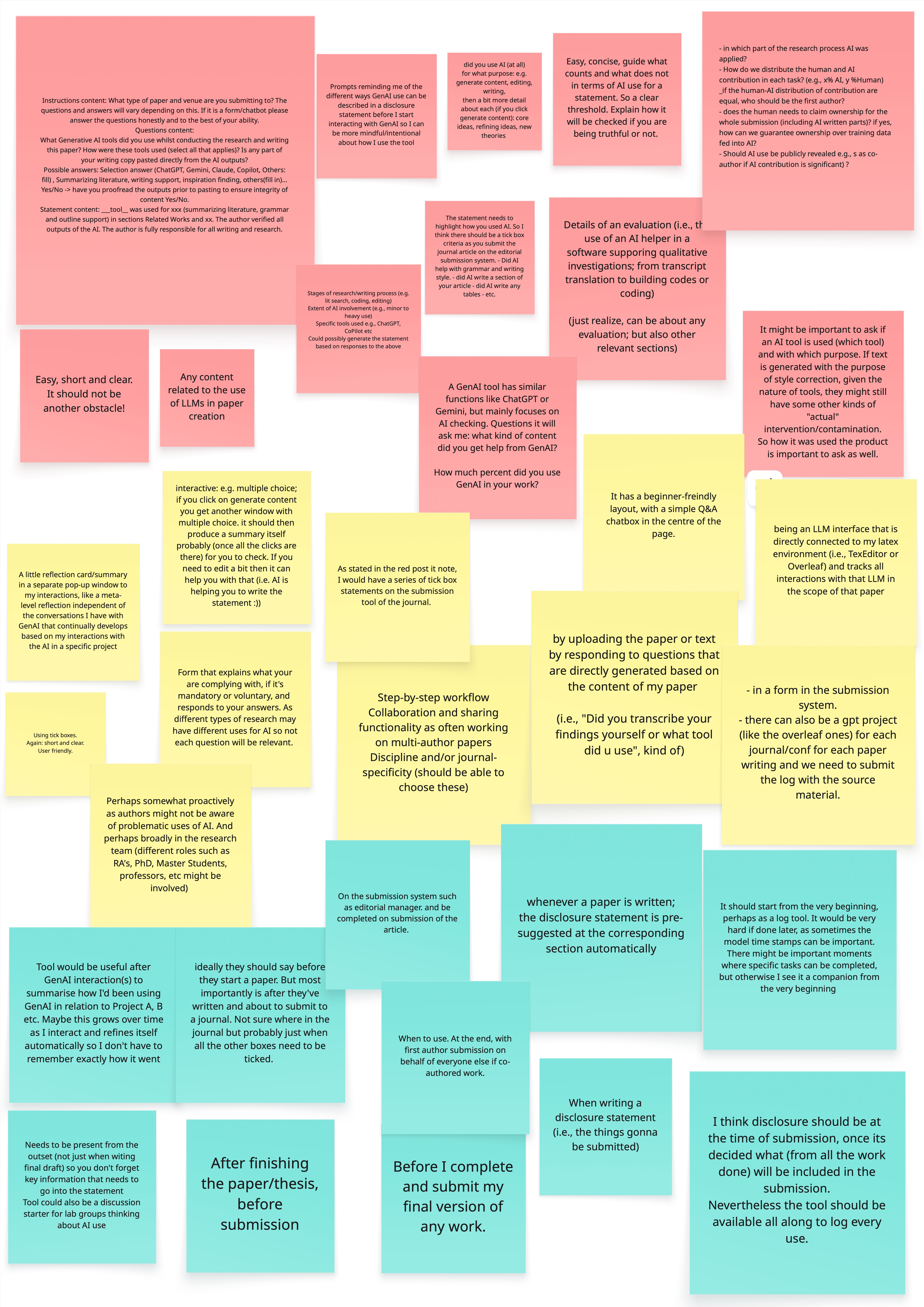} \caption{Post-It Notes Created During the Co-Design Sessions} 
    \label{fig:postits} 
\end{figure*} 

\newpage
\section{Thematic Analysis Used for Research Group Discussions and Refinement} 
\label{sec:appendix-thematic}
    \begin{figure*}[h!]
    \centering 
    \includegraphics[width=1\linewidth]{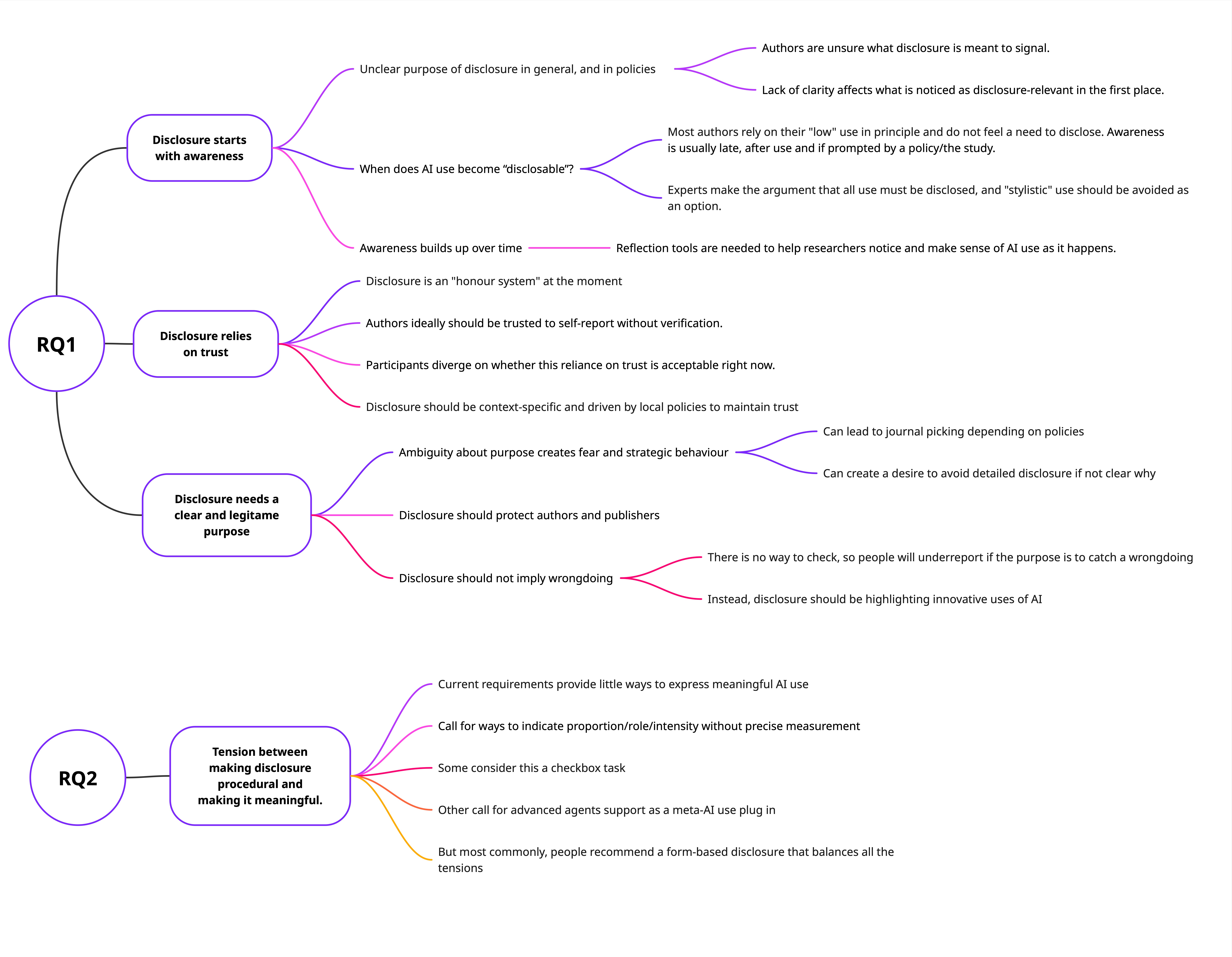} \caption{Thematic Analysis Used for Research Group Discussions and Refinement} 
    \label{fig:map} 
\end{figure*}

\end{document}